\documentclass[a4paper,twoside]{article}
\newcommand{\affil}[1]{$^{\rm #1}$}
\usepackage[authoryear]{natbib}
\bibpunct{(}{)}{;}{a}{}{,}
\usepackage{graphicx}

\usepackage{subfigure}
\usepackage{rotating}
\usepackage{color}
\definecolor{grey}{rgb}{0.5,0.5,0.5}
\usepackage{amsmath}
\usepackage{amstext}
\usepackage{amssymb}

\date{} 
\newcommand{\aap}{A\&A}   
\newcommand{\aaps}{Astron. Astrophys. Suppl. Ser.}
\newcommand{\ao}{Appl. Opt.}
\newcommand{\josa}{J. Opt. Soc. Am.}
\newcommand{\josaa}{J. Opt. Soc. Am. A}
\newcommand{\optcomm}{Opt. Commun.}
\newcommand{\pasp}{PASP}
\newcommand{\mnras}{MNRAS}
\newcommand{\spie}{Proc. SPIE}

\newcommand{\drunits}{\ensuremath{\mathrm{m}.\mathrm{pix}^{-1}}}
\newcommand{\vhunits}{\ensuremath{\mathrm{m} \mathrm{s}^{-1}}}
\newcommand{\cn}{\ensuremath{C_N^2(h)}}
\newcommand{\vw}{\ensuremath{V(h)}}
\newcommand{\vwi}{\ensuremath{V(h_i)}}
\newcommand{\s}{\space}
\newcommand{\fg}{\ensuremath{f_G}}
\newcommand{\mjuo}{Mount John University Observatory}
\newcommand{\dt}{\ensuremath{\Delta t}}
\newcommand{\dr}{\ensuremath{\Delta r}}
\newcommand{\dm}{\ensuremath{\Delta m}}
\newcommand{\pup}{pupil-plane}
\newcommand{\gen}{generalised}
\newcommand{\Pup}{Pupil-plane}
\newcommand{\Gen}{Generalised}
\newcommand{\kolm}{Kolmogorov}
\newcommand{\acen}{$\alpha$ Cen}
\newcommand{\acru}{$\alpha$ Cru}
\newcommand{\degree}{\ensuremath{^\circ}}
\renewcommand{\dh}{\ensuremath{\Delta h}}
\newcommand{\cndhunits}{\ensuremath{\mathrm{m}^{1/3}}}
\newcommand{\averzero}{\ensuremath{\overline{r_0}}}
\newcommand{\avethetazero}{\ensuremath{\overline{\theta_0}}}

\newcommand{\stdthetazero}{\ensuremath{\sigma_{\theta_0}}}
\newcommand{\avefg}{\ensuremath{\overline{\fg}}}
\newcommand{\stdfg}{\ensuremath{\sigma_{\fg}}}
\newcommand{\vwmag}{\ensuremath{|\vw|}}
\newcommand{\vh}{\vw}
\newcommand{\cnunits}{\ensuremath{\mathrm{m}^{-2/3}}}
\title{\large\bf\flushleft Optical Turbulence Measurements and Models for Mount John University Observatory}
\author{\parbox{\textwidth}{\flushleft
\vspace{-0.5cm}
{\it J. L. Mohr\affil{A,C}, R. A. Johnston\affil{B}, and P. L. Cottrell\affil{A}}\\
\vspace{0.4cm}
{\small \affil{A}\,Department of Physics and Astronomy, University of Canterbury, Private Bag 4800, Christchurch 8020, New Zealand}\\
{\small \affil{B}\,ARANZ Scanning Ltd. (ASL), PO Box 3894, Christchurch 8011, New Zealand}\\
{\small \affil{C}\,Email: j.mohr@phys.canterbury.ac.nz}}}
\begin{document}
\twocolumn[
\begin{changemargin}{.8cm}{.5cm}
\begin{minipage}{.9\textwidth}
\vspace{-1cm}
\maketitle
%
%
\small{\bf Abstract:} Site measurements were collected at Mount John University Observatory in 2005 and 2007 using a purpose-built scintillation detection and ranging system.  \cn\s profiling indicates a weak layer located at 12 -- 14 km above sea level and strong low altitude turbulence extending up to 5 km.  During calm weather conditions, an additional layer was detected at 6 -- 8 km above sea level. \vw\s profiling suggests that tropopause layer velocities are nominally 12 -- 30 \vhunits, and near-ground velocities range between 2 -- 20 \vhunits, dependent on weather.  Little seasonal variation was detected in either \cn\s and \vw\s profiles.  The average coherence length, $r_0$, was found to be $7 \pm 1$ cm for the full profile at a wavelength of 589 nm.  The average isoplanatic angle, $\theta_0$, was $1.0 \pm 0.1$ arcsec.  The mean turbulence altitude, $\overline{h_0}$, was found to be $2.0\pm0.7$ km above sea level.  No average in the Greenwood frequency, \fg, could be established due to the gaps present in the \vw\s profiles obtained.  A modified Hufnagel-Valley model was developed to describe the \cn\s profiles at Mount John, which estimates $r_0$ at 6 cm and $\theta_0$ at 0.9 arcsec.  A series of \vw\s models were developed, based on the Greenwood wind model with an additional peak located at low altitudes.  Using the \cn\s model and the suggested \vw\s model for moderate ground wind speeds, \fg\s is estimated at 79 Hz.



\medskip
\medskip
\end{minipage}
\end{changemargin}
]
\small

\section{Introduction}

Astronomical images taken by ground based telescopes are subject to distortion caused by atmospheric turbulence.  Adaptive optics (AO) provides a real-time solution to compensate for an aberrated wavefront through the use of deformable optics in a closed-loop system.  Accurate measurements and models of atmospheric turbulence are an essential tool in the design and optimisation of an AO system \citep{AvilaJ65}.

Key parameters for the design of an AO system include the turbulence coherence length, $r_0$, the isoplanatic angle, $\theta_0$, and the Greenwood frequency, \fg.  $r_0$ describes the effective telescope diameter for which nearly diffraction-limited resolution can be obtained if no attempt is made to compensate for atmospheric turbulence.  It is defined as \citep{TysonB42}
\begin{eqnarray}\label{eqn:theory:r0}
    r_0 = \left[ 0.423 k^2 \sec(\zeta) \int \cn \mathrm{d}h \right]^{-3/5},
\end{eqnarray}
where $k = 2\pi /\lambda$ is the wavenumber for a given wavelength
$\lambda$ and $\zeta$ is the zenith angle.  \cn\s is the refractive index structure constant and is a measure of the strength of a turbulent layer located at altitude $h$.

The isoplanatic angle, $\theta_0$, describes the maximum angular separation between two objects for which turbulence induced distortions are essentially identical  and is defined as \citep{ParentiJ100}
\begin{eqnarray}
    \theta_0 = \left[ 2.91 k^2 \sec^{8/3}(\zeta) \int \cn
    h^{5/3} \mathrm{d}h \right]^{-3/5}.
    \label{eqn:theta0}
\end{eqnarray}
Unlike $r_0$, $\theta_0$ is dependent on $h^{5/3}$ indicating that weak high altitude layers have a significant impact on $\theta_0$.

Atmospheric turbulence is in a constant state of motion.  The Greenwood frequency, $f_G$,  describes the rate at which the turbulence structure above a site changes with time.
 It is defined as \citep{TysonB48}
\begin{equation}
f_G = 0.255 \left[ k^2 \sec \zeta \int
\cn \vw^{5/3} \mathrm{d}h \right]^{3/5} \label{eqn:theory:Greenwood},
\end{equation}
where \vw \s is the average horizontal wind velocity as a function of altitude $h$.  $f_G$ determines how quickly an AO system must respond to adequately compensate for the aberrations induced by atmospheric turbulence.

The 1-m McLellan telescope at \mjuo\s (MJUO), located at Tekapo, New Zealand, is used for a variety of different astronomical research and is known to regularly experience poor seeing ($>2$ arcsec angular resolution) by observers (A. Gilmore (MJUO) 2006, private communication).  This work is part of a feasibility study on installing an AO system to improve photometric images with the CCD photometer head and to improve light throughput into the HERCULES \'{e}chelle spectrograph \citep{HearnshawExA2002} currently installed on the 1-m telescope.  The elevation of MJUO is 1024 m.

This paper discusses the \cn\s and \vw\s profile measurements taken at MJUO and the models developed for use in an AO system design.  Section \ref{sec:measuring} outlines the techniques used to measure \cn\s and \vw\s profiles and the purpose-built system.  Sections \ref{sec:data} and \ref{sec:trends} discuss the data collected and the trends noted in the profiles obtained.  Section \ref{sec:model} introduces the models developed to describe atmospheric turbulence at MJUO.  Concluding remarks are in section \ref{sec:conclusions}.

\section{Measuring Turbulence}
\label{sec:measuring}
SCIDAR (SCIntillation Detection And Ranging) is a remote sensing technique that has been used at many different sites around the world to characterise optical turbulence \citep{AvilaJ149, GarciaLorenzoMNRAS2009, MasciadriMNRAS2010, PrieurJ58, TokovininJ333}.  It uses the spatio-temporal covariance functions obtained from a sequence of short exposure images of the scintillation pattern seen at a telescope pupil to infer the \cn\s and \vw\s profiles present above a site \citep{KluckersJ67}.

\begin{figure}
  \centering
  \includegraphics[width=0.95\linewidth, trim=10mm 10mm 10mm 150mm]{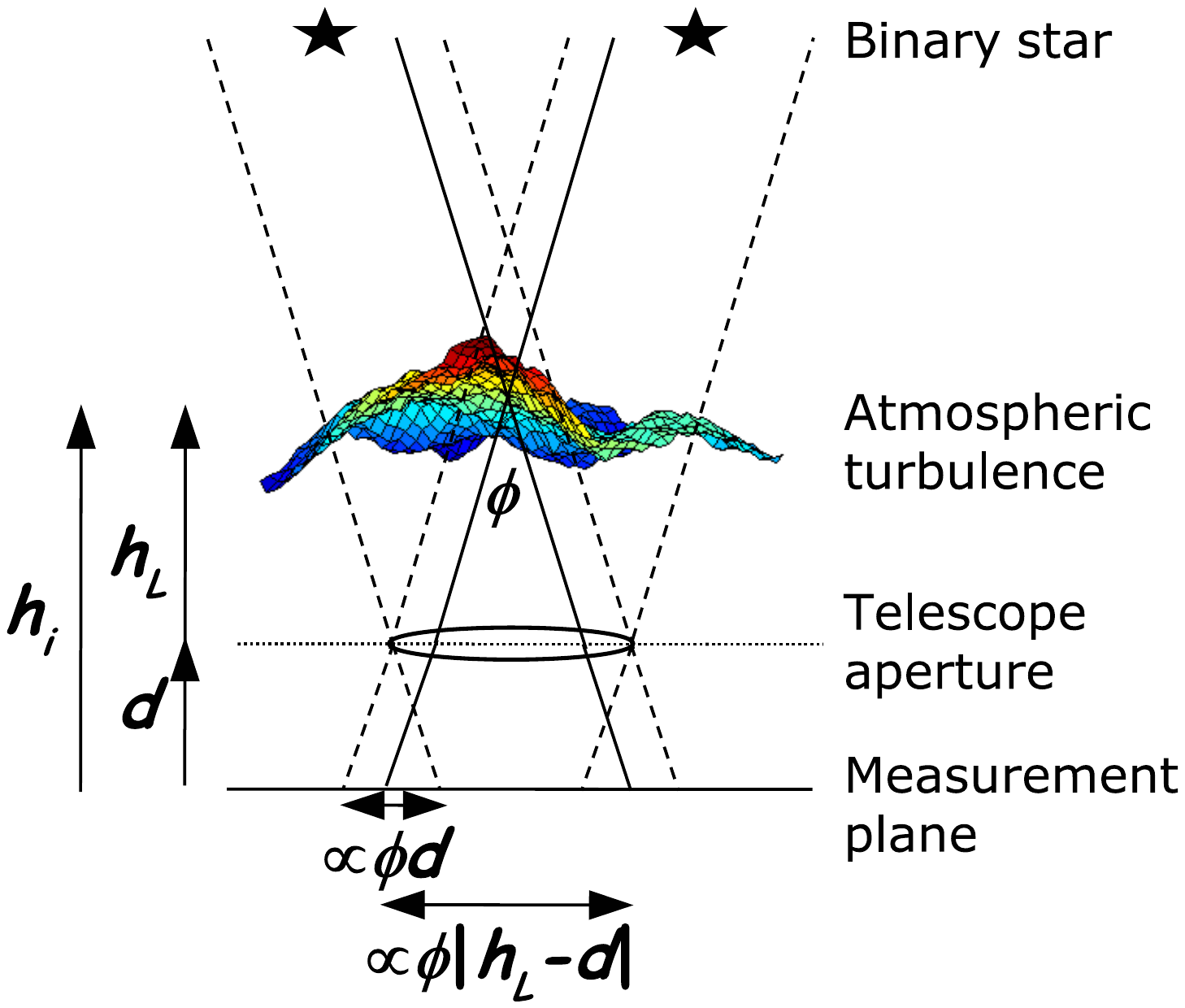}
  \caption{The concept of double star SCIDAR.  Light from each star passes through the same turbulent region forming identical scintillation patterns separated by a distance proportional to the double star separation $\phi$ and the height of the turbulent layer above the measurement plane $h_i = |h_L - d|$.  Due to sign conventions $d$ is negative.}\label{fig:theory:scidarConcept}
\end{figure}

SCIDAR measurements are commonly taken using a double star system, as indicated in Figure \ref{fig:theory:scidarConcept}.  Light from each star
passes through the same region of a turbulent layer forming
identical, but separated, scintillation patterns.  The distance between the
two scintillation patterns is directly proportional to the angular
separation of the double star system, $\phi$, and the height of the
turbulent layer above the measurement plane, $h_i$.

\emph{Pupil-plane} SCIDAR measures scintillation \linebreak patterns seen at the telescope aperture.  In doing this, a clear picture of optical turbulence in the free atmosphere can be obtained.  As scintillation is proportional to $h^{5/6}$ \citep{RoddierAB12}, any scintillation resulting from near-ground turbulence (NGT) is not readily detectable.  Using a simple lens change, the measurement plane can be shifted to a virtual plane located at $d$ below the telescope.  If $h_L$ is the height of the layer above the telescope then the height of the layer above the measurement plane becomes $h_i=|h_L - d|$, where $d$ is negative due to sign conventions.  This increased propagation distance allows for scintillation from NGT to be adequately measured.  This version of SCIDAR is known as \emph{generalised} SCIDAR \citep{KluckersJ67}.

The University of Canterbury SCIDAR system \linebreak(UC-SCIDAR) is a purpose-built instrument designed to measure \cn\s and \vw\s profiles (Johnston et al. 2004, 2005; Mohr at al. 2006, 2008a, 2008b). 
UC-SCIDAR saw first light on the McLellan 1-m telescope at MJUO in late 2003, at an approximate cost of USD\$4000.  To keep costs low the system design utilised primarily
off-the-shelf components.  Over the years, the system has evolved through several iterations.

\begin{figure}
\centering
\subfigure[Physical layout]{\includegraphics[width=0.5\linewidth, trim=0mm 0mm 0mm 150mm]{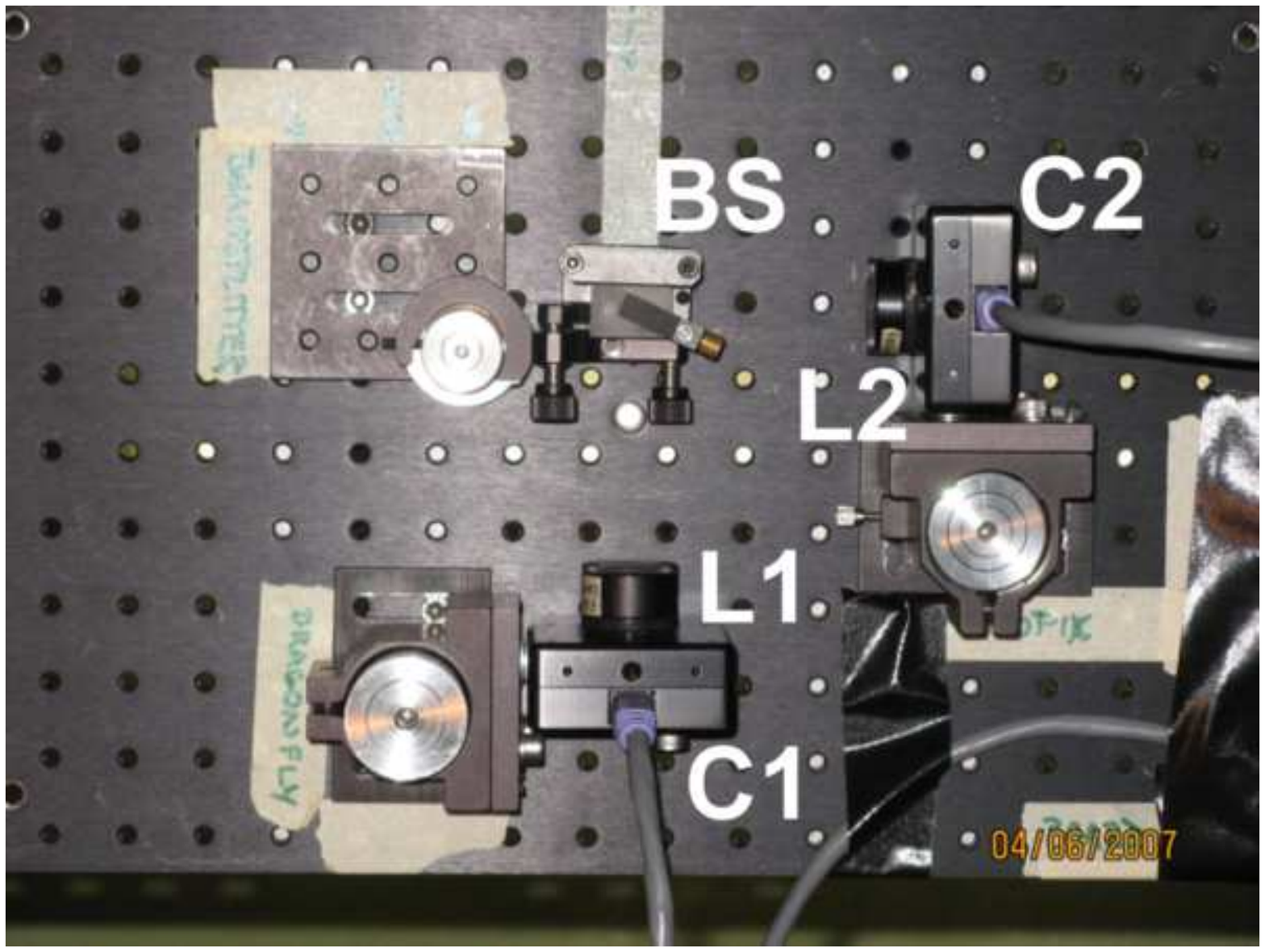}\label{fig:rig:v2007}}
\hspace{0.1\linewidth}
\subfigure[Optical layout]{\includegraphics[width=0.35\linewidth, trim=0mm 0mm 80mm 190mm]{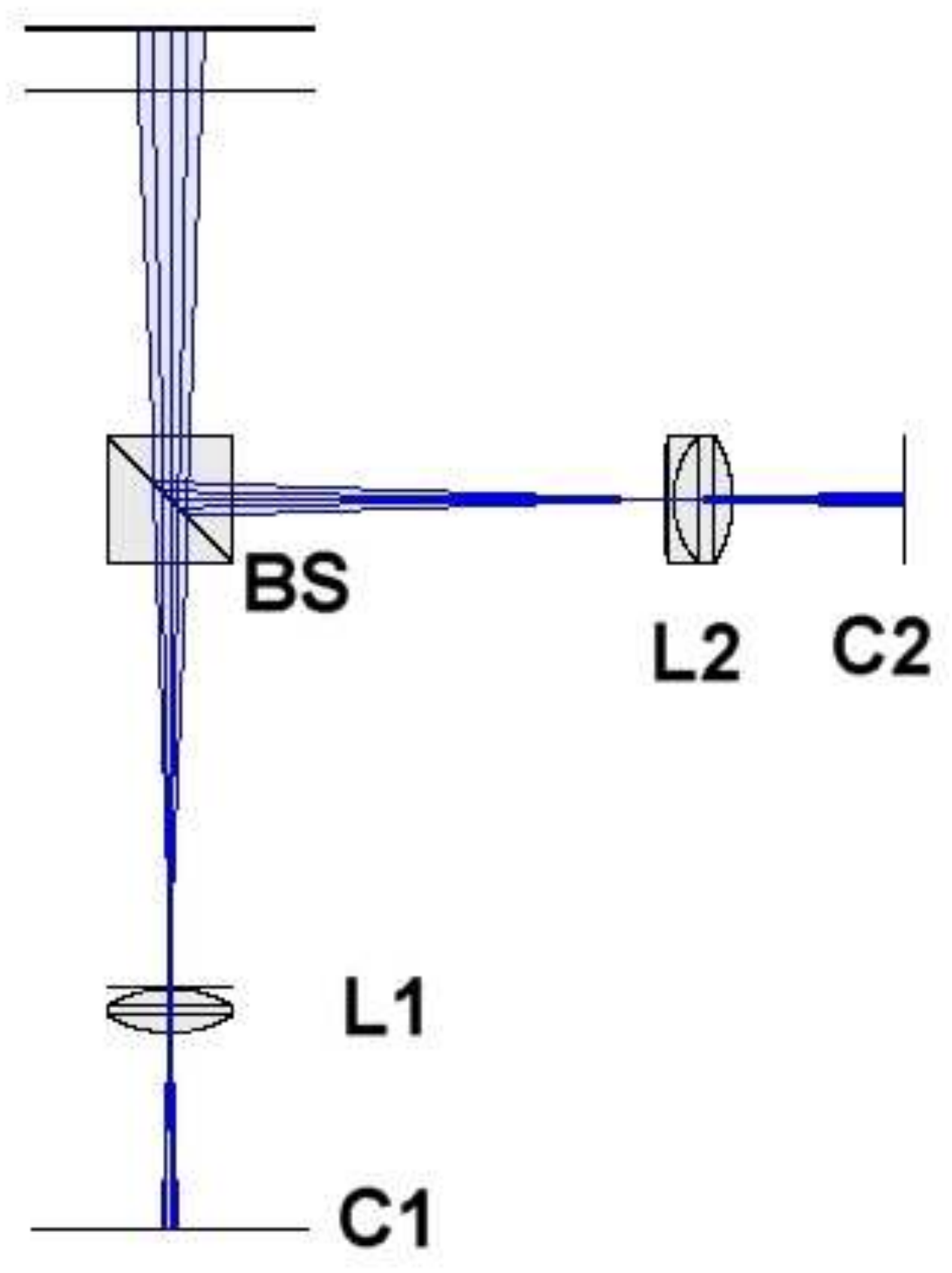}\label{fig:rig:optical}}
  \vspace{-6pt}
\caption{\subref{fig:rig:v2007} Physical and \subref{fig:rig:optical} optical layout of the UC-SCIDAR instrument.  See Section \ref{sec:measuring} for a detailed description of the system.} \label{fig:rig}
\vspace{6pt}\end{figure}

The current system consists of two channels, each with its own CCD camera and field  lens (Figure \ref{fig:rig}).  Light from the telescope is split using a 50/50 intensity beamsplitter (BS) to minimize the differences between the two channels.  The straight path, typically used for pupil-plane SCIDAR, consists of a f12.7mm achromat lens (L1) mounted in a lens tube a focal length away from a Dragonfly Express CCD camera from Point Grey Research (C1).  This camera uses a Kodak KAI-0340DM sensor which has a 640 x 480 grid of 7.4 $\mu$m square pixels. UC-SCIDAR has been configured to capture images at full resolution with a frame rate of 60 Hz.  Operational exposure times range from 0.5 to 5 ms.  The lens used provides a nominal spatial sampling, \dr, of $1/125$ m.pix$^{-1}$ when the 1-m telescope is operating at a focal ratio of $F/13.5$.

\begin{figure}
\centering
\subfigure[\Pup\s frame]{\includegraphics[width=0.45\linewidth, trim=0mm 50mm 0mm 90mm]{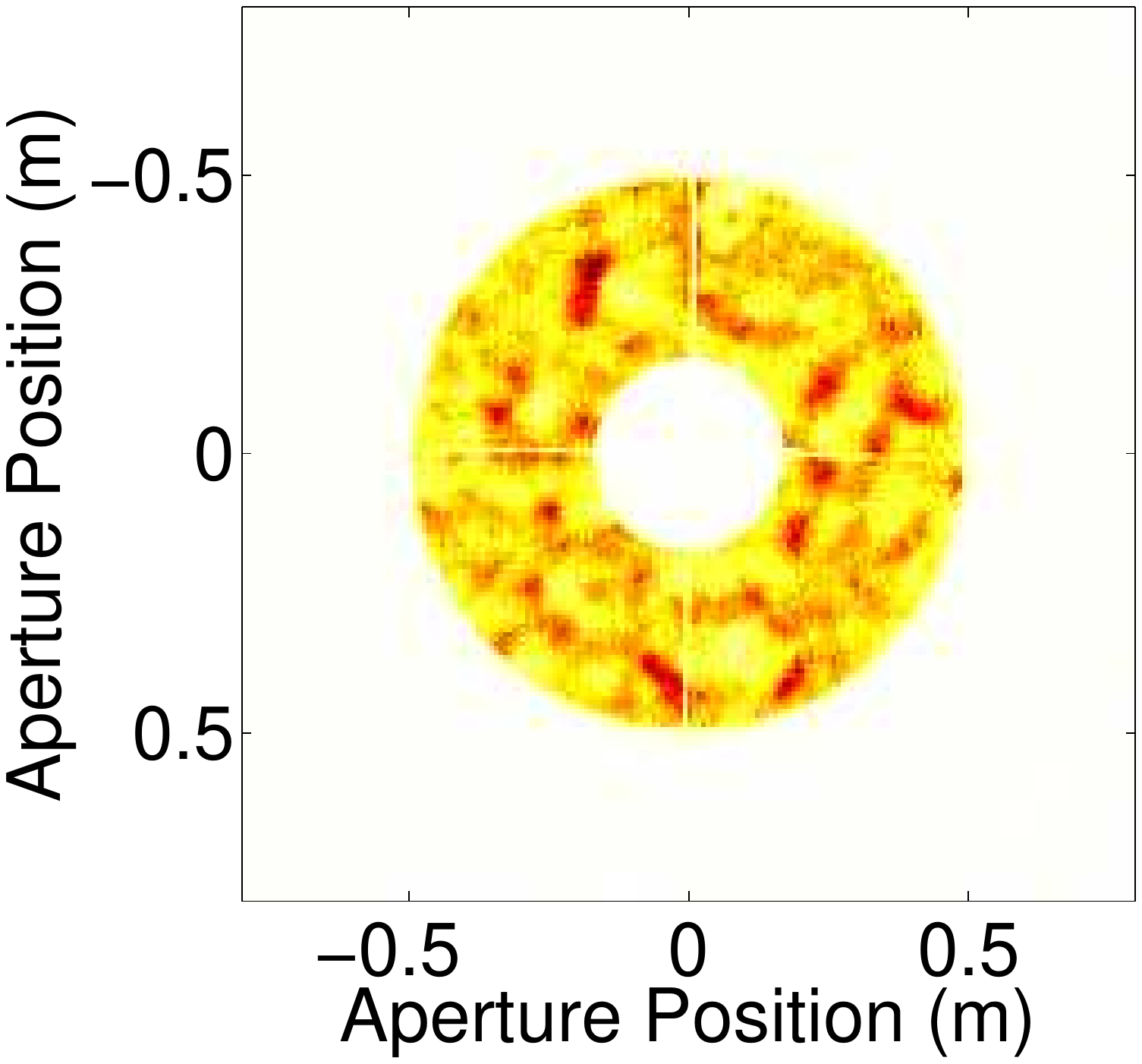}\label{fig:spatial:scintFrames:realPup}}
\subfigure[\Gen\s frame]{\includegraphics[width=0.45\linewidth, trim=0mm 50mm 0mm 90mm]{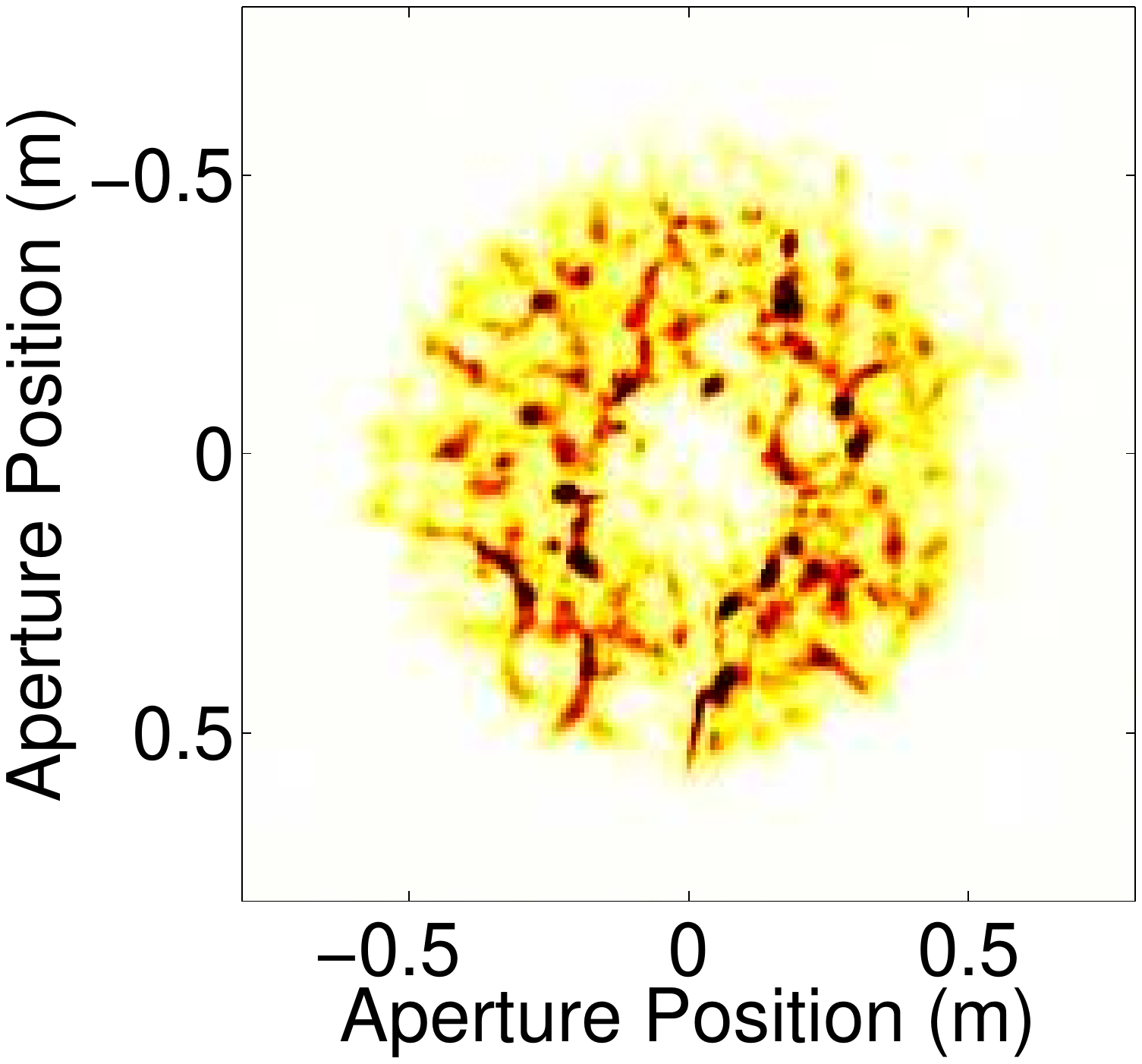}\label{fig:spatial:scintFrames:realGen}}
  \caption[Typical scintillation images from \pup\s and \gen\s SCIDAR.]{Typical scintillation images from \subref{fig:spatial:scintFrames:realPup} \pup\s and \subref{fig:spatial:scintFrames:realGen} \gen\s SCIDAR.
}\label{fig:scidar:typicalImages}
\end{figure}

A second identical CCD camera is mounted in the side path (C2).  As this channel is typically used for generalised SCIDAR, a f10mm achromat lens (L2) is mounted a focal length of L1 away from C2.  This provides a measurement plane at approximately 3.5 km below the telescope.  Figure \ref{fig:scidar:typicalImages} shows a typical \pup\s and \gen\s SCIDAR scintillation image obtained using UC-SCIDAR.

Assuming that exposure times and frame rates are short enough that turbulent elements move without distortion, then the 2D spatio-temporal covariance \linebreak function for a double star can be written as \citep{AvilaJ65}
\begin{eqnarray}
    C_B(\rho,\phi,\dt) & = &\sum_{i=1}^n \{aC_S(\rho - \vwi\dt)  \nonumber\\
        & & + bC_S(\rho - \vwi\dt - \phi h_i) \nonumber\\
        &&  + bC_S(\rho - \vwi\dt + \phi h_i)\},\label{eqn:tempT:covariance_binary}
\end{eqnarray}
where $\phi$ is the angular separation of the double star, \dt\s is the time delay between consecutive scintillation images, $n$ is the number of discrete turbulent layers present and $C_S(\rho - \vwi\dt)$ is the spatio-temporal covariance of a single star for the radial coordinate, $\rho$, in the direction of the double star.  The coefficients $a$ and $b$ are
given by \citep{AvilaJ65}
\begin{equation}\label{eqn:tempT:corrFxnCoeffs_binary}
    a = \frac{1+\alpha^2}{(1+\alpha)^2}, b =
    \frac{\alpha}{(1+\alpha)^2}, \alpha = 10^{-0.4\dm},
\end{equation}
where \dm\s is the magnitude difference between the double star components.  $C_B(\rho,\phi,\dt)$ describes a series of triplets where each corresponds to a different turbulent layer.  

When $\dt = 0$ in equation (\ref{eqn:tempT:covariance_binary}), the secondary peak, $C_S(\rho - \phi h_i)$, contains all the required information to determine the altitude and \cn\s strength for a given layer.  It has been shown that $C_S(\rho - \phi h_i)$ is approximately the difference between parallel and perpendicular slices from the 2D covariance with respect to the direction of the double star, $C_{B,\|}$ and $C_{B,\bot}$ respectively, and can be written as \citep{AvilaJ138}
\begin{eqnarray}\label{eqn:secondaryCn2}
    C_S(\rho - \phi h_i) & \approx & C_{B,\|}(\rho, \phi, 0) - C_{B,\bot}(\rho, \phi, 0) \\
    & = &\int_0^\infty K(\rho, h_i)C_N^2(h_i)\mathrm{d}{h_i} + n(\rho),\nonumber
\end{eqnarray}
where $n(\rho)$ is the measurement noise, resulting in a single covariance peak shifted by $\phi h_i$. (Note the change in subscripts.)  $K(\rho, h_i)$ is the theoretical spatial covariance of the scintillation from a single star produced by a layer at height $h_i$ where $\int C_N^2(h_i)\mathrm{d}{h_i} = 1$, assuming \kolm\s turbulence.

The altitude of a given later can be found from the spatial sampling across the aperture, \dr, and the stellar separation, $\phi$, where the altitude resolution of the system, \dh, can be defined as \citep{KluckersJ67}
\begin{equation}\label{eqn:theory:dhDefinition}
    \dh = \frac{\dr}{\phi\sec\zeta}.
\end{equation}
$\zeta$ is the angle from zenith that the measurement was taken at.  For $\dr = 0.01$ \drunits\s and $\phi = 4$ arcseconds, $\dh \approx 500$ \drunits\s at zenith.

Using the methods discussed in \citet{JohnstonJ200} and \citet{JohnstonJ44}, \cn\s profiles can be obtained from UC-SCIDAR data.

Using $\dt > 0$, it is assumed that if a layer moves with a horizontal wind velocity \vwi, then the scintillation pattern produced at the ground would move with the same \vwi.  For a single turbulent layer, two scintillation patterns separated in time by \dt\s would be separated by \vwi\dt\s \citep{AvilaJ65}. Hence
\begin{equation}\label{eqn:tempT:TaylorHypothesis}
    C_B(\vwi\dt,\phi,\dt) \approx C_B(0,\phi,0),
\end{equation}
such that the height of the auto-covariance peak (i.e. $\dt = 0$) is approximately the
height of the spatio-temporal cross-covariance peak displaced by a
distance \vwi\dt.  To adequately capture motion of slowly moving layers a long \dt\s is required.  However, long \dt\s values are blind to rapidly moving layers.  Using multiple values of \dt, a full \vw\s profile can be obtained.  The algorithm used to analyse spatio-temporal covariances for \vw\s profiles is presented in \citet{MohrIVCNZ08}.

%

\section{Data For Trending}
\label{sec:data}

The first SCIDAR measurements taken at MJUO occurred in April 1999 using the system designed by Imperial College \citep{JohnstonJ44}.  These data revealed the presence of strong NGT  and two different high altitude layers located at approximately 11 and 13 km above sea level, with an estimated $r_0$ of 12.3 cm for the full profile.  Temporal analysis indicated that the velocities of the 11 and 13 km layers were 6.45 and 11.63 \vhunits\s respectively, whereas the NGT layer was attributed to dome seeing.  However the measurements were collected over a single observation run of 10 nights where only 50\% of the nights provided useful observing conditions \citep{JohnstonP2}.  For the development of AO for MJUO, it was decided that a more complete picture was required of variations in the turbulence profile with respect to season and weather.

SCIDAR measurements for trending purposes were collected at MJUO from 2005 -- 2007 using \linebreak UC-SCIDAR.  The majority of measurements were \linebreak taken using the 1-m McLellan telescope at a focal ratio of $F/13.5$.  Due to the noise characteristics of the CCD cameras only a handful of double or binary star systems were suitable.  The location of MJUO permitted the binary star systems \acru\s and \acen\s to be used for a large portion of the year.  However in the summer months, when both \acru\s and \acen\s are too far from zenith, fainter star systems such as $\theta$ Eri and $\upsilon$ Car could be used with the latest version of UC-SCIDAR, which has more sensitive detectors.  For the UC-SCIDAR system stellar separations should be limited to between 4 and 20 arcsec with the apparent magnitude of the primary star no fainter than 3.5, where the magnitude difference should be limited to $\sim 2.5$.  Table \ref{tab:stellarParam} gives the stellar parameters for the stars for measurements presented in this paper.

\begin{table}
  \centering
  \caption{Stellar parameters for stars used with UC-SCIDAR data presented.}
  \label{tab:stellarParam}
  \begin{center}\begin{tabular}{cccc}
\hline
 Star &  $\phi$ (arcsec) &  $m_1$ & $\Delta m$ \\
\hline
 \acen &  8.7 -- 13.3 &  -0.01 & 1.36 \\
 \acru &  3.9 &  1.25 & 0.3 \\
 $\theta$ Eri &  8.4 &  3.20 & 0.92 \\
 $\upsilon$ Car &  5.0 &  3.02 & 2.98 \\
\hline
\end{tabular}\end{center}
\end{table}

Not all of the data collected between 2005 -- 2007 was suitable for use in the profiling of turbulence at MJUO.  Only data with exposure times $\leq 3.5$ ms were included in the trending analysis, to ensure that turbulence was not subject to excessive blurring, resulting in the underestimation of layer strengths.  \citet{KluckersJ67} used exposure times raging from 1.6 -- 2.7 ms, whereas \citet{AvilaMNRAS2008} report use of exposures of 3 ms.  Measurements taken at zenith angles, $\zeta$, greater than 35\degree\s were excluded from site profiling.  Measurements with $30\degree < \zeta \leq 35\degree$ were included only if supporting data taken at $\zeta \leq 30\degree$ was acquired within approximately 30 minutes of the run.  This was to ensure that a suitable altitude range was sampled and that measured NGT was not subject to vertical air flows located near ground.  A significant portion of the data collected during 2006 was subject to corrupt CCD readout and hence was not included in this study.

Each observation period usually consisted of three to four consecutive nights.  Data sequences typically consisted of 5000 frames from each camera which was recorded into file blocks of 500 frames per file.  To decrease the processing time, 2D spatio-temporal covariances for each camera were calculated using 2500 frames with time delays between consecutive frames of $\dt = 0, \mathrm{d}t, 2\mathrm{d}t \ldots 6\mathrm{d}t$, where $\mathrm{d}t$ represents the frame rate of the CCD camera used.  This ensured that at least 1000 cross-correlations were used in the longer \dt\s ensembles.  It should be noted that all available frames were used such that for a $\dt = 3\mathrm{d}t$, frame 1 was correlated with frame 4, frame 2 with frame 5, frame 3 with frame 6, and so on.

Measured velocities were cross-checked between sequential runs and the various \dt\s covariances, and layer heights from temporal analysis were cross-checked with the corresponding \cn\s analysis for a given run to eliminate any falsely detected layers.  Data collected in 2007 was collected at 60 Hz, whereas data collected prior to 2007 was collected at 30 Hz due to a limitation in the CCD cameras used at the time.

In a significant number of cases there was simultaneous \pup\s and \gen\s SCIDAR data.  This provided an added check to layer heights in the free atmosphere, as well as an insight into how the strong low altitude layers can affect the measurements obtained for high altitude layers.

Across the three years, a total of 324 \pup\s and 315 \gen\s SCIDAR data sequences were suitable for \cn\s profiling.  134 \pup\s and 112 \gen\s SCIDAR data sequences were used for \vw\s profiling.  The significant decrease in the number of suitable runs for \vw\s profiling was due to the frame rate of early UC-SCIDAR data and the nature of NGT present at the time in later UC-SCIDAR data.  Strong NGT can blur any covariance peaks detected during temporal analysis, particularly when using star systems with a narrow angular separation.  As such it is not always possible to determine the height of layers using temporal analysis as the primary and secondary peaks blur together.

The \cn\s profiles presented in this paper include dome/mirror seeing as any AO system developed for the McLellan 1-m telescope at MJUO would need to compensate for dome/mirror seeing too.

All $r_0$, $\theta_0$ and \fg\s values calculated from \linebreak UC-SCIDAR data have been determined for a wavelength, $\lambda$, of 589 nm, although the measurements were collected using broadband \emph{white} light.  The variable weather at MJUO (ranging from calm, clear nights to gusting winds and thickening clouds) resulted in a variety of profiles being detected.  During increasing cloud cover, with moderate to high ground wind speeds, the NGT present was exceptionally strong resulting in a \pup\s $r_0$ that was similar to that found for the corresponding \gen\s data during analysis.  In these cases the \gen\s $r_0$ estimate was used in the determination of the \pup\s averages to remove any bias toward noise in \pup\s results.  Results are presented from the most recent to the earliest.

\section{Trends at MJUO}
\label{sec:trends}

\subsection{\cn\s Profiles}

\begin{figure}
  \centering
  \subfigure[\Pup]{\includegraphics[width=\linewidth, trim=0mm 0mm 0mm 190mm]{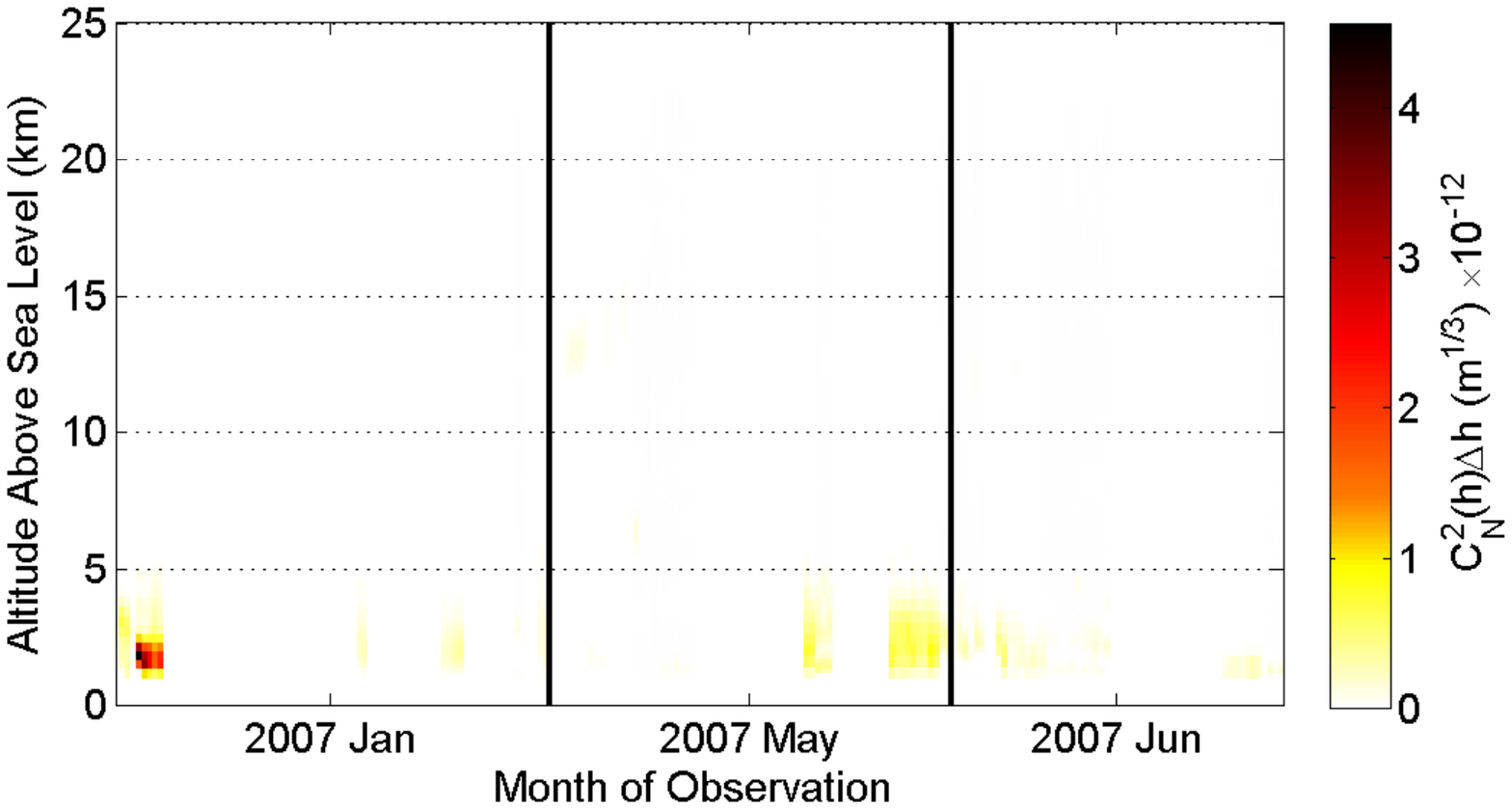}\label{fig:trending:2007_cn2P:norm}}
  \subfigure[Scaled \Pup]{\includegraphics[width=\linewidth, trim=0mm 0mm 0mm 190mm]{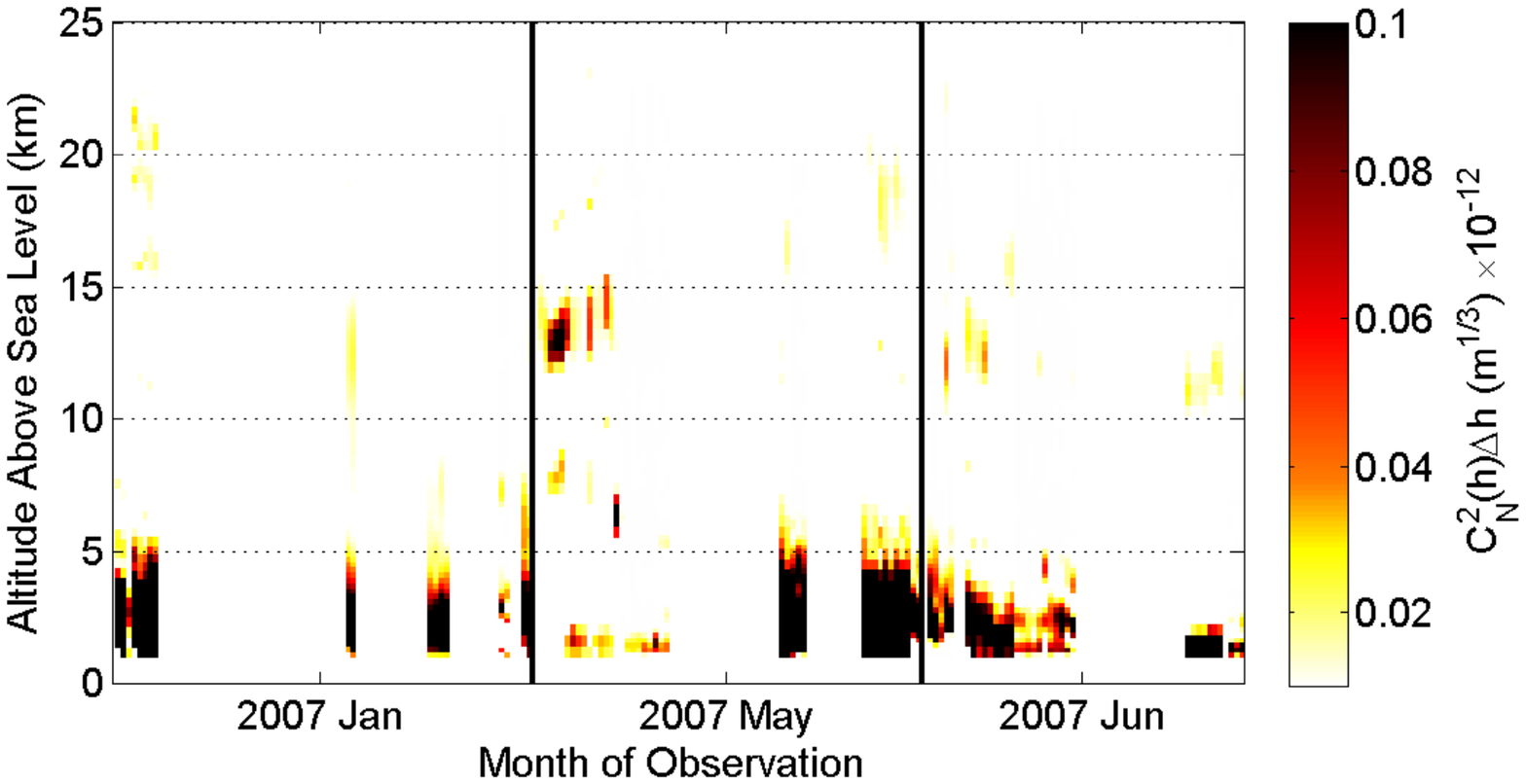}\label{fig:trending:2007_cn2P:scaled}}
  \caption[\Pup\s \cn\s profile trends observed over 2007.]{\Pup\s \cn\s profile trends observed over 2007. Gaps have been added where no data is present for more than two hours.  Data from the individual months is separated by solid black lines.  The image in \subref{fig:trending:2007_cn2P:scaled} has been scaled such that any \cn\dh\s value above $10^{-13} \mathrm{m}^{1/3}$ is set to the maximum colour range and any \cn\dh\s value below $10^{-14} \mathrm{m}^{1/3}$ is set to the minimum colour range.}\label{fig:trending:2007_cn2P}
\end{figure}

\begin{figure}
  \centering
  \subfigure[\Gen]{\includegraphics[width=\linewidth, trim=0mm 0mm 0mm 190mm]{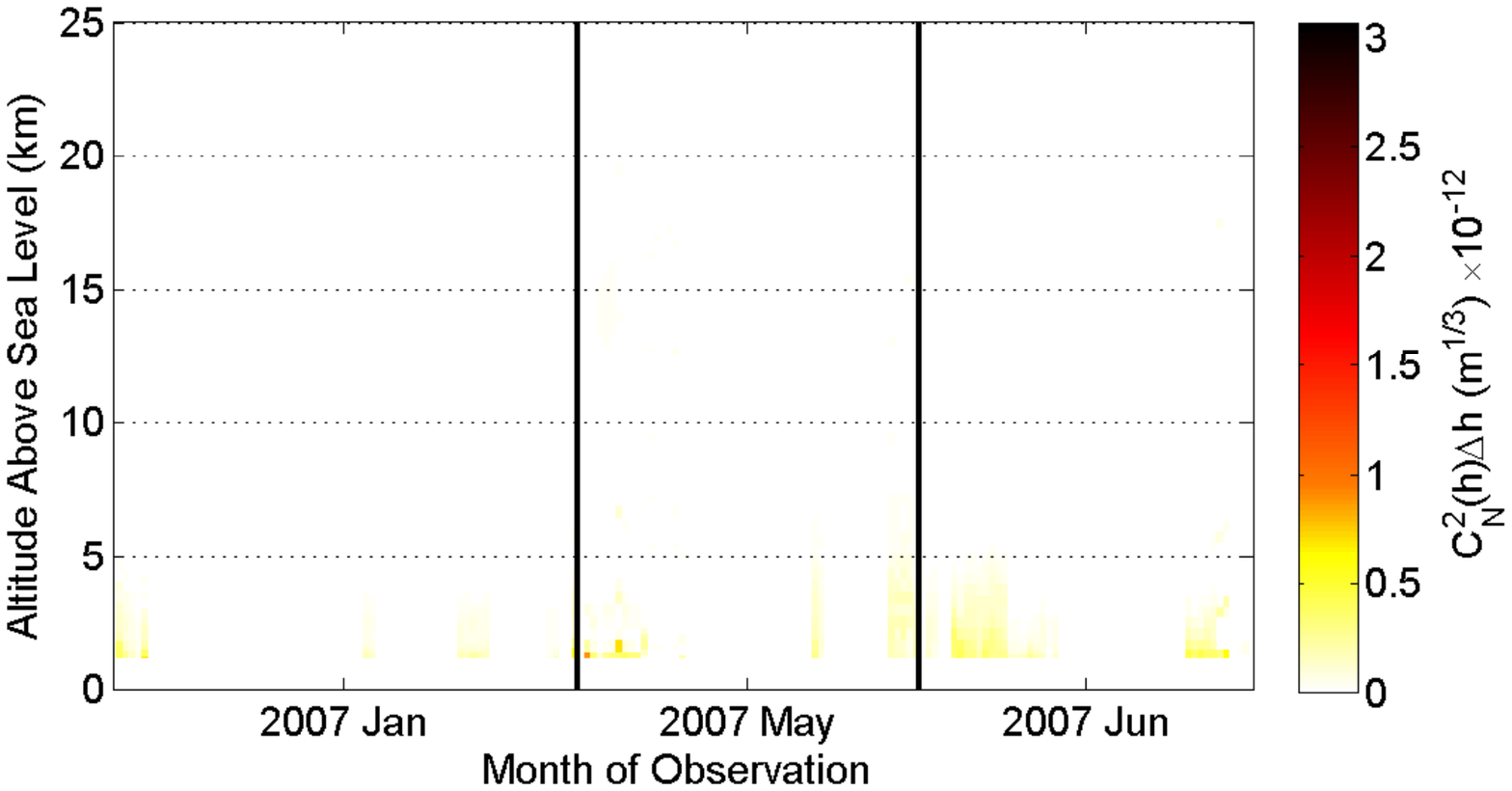}\label{fig:trending:2007_cn2G:norm}}
  \subfigure[Scaled \Gen]{\includegraphics[width=\linewidth, trim=0mm 0mm 0mm 190mm]{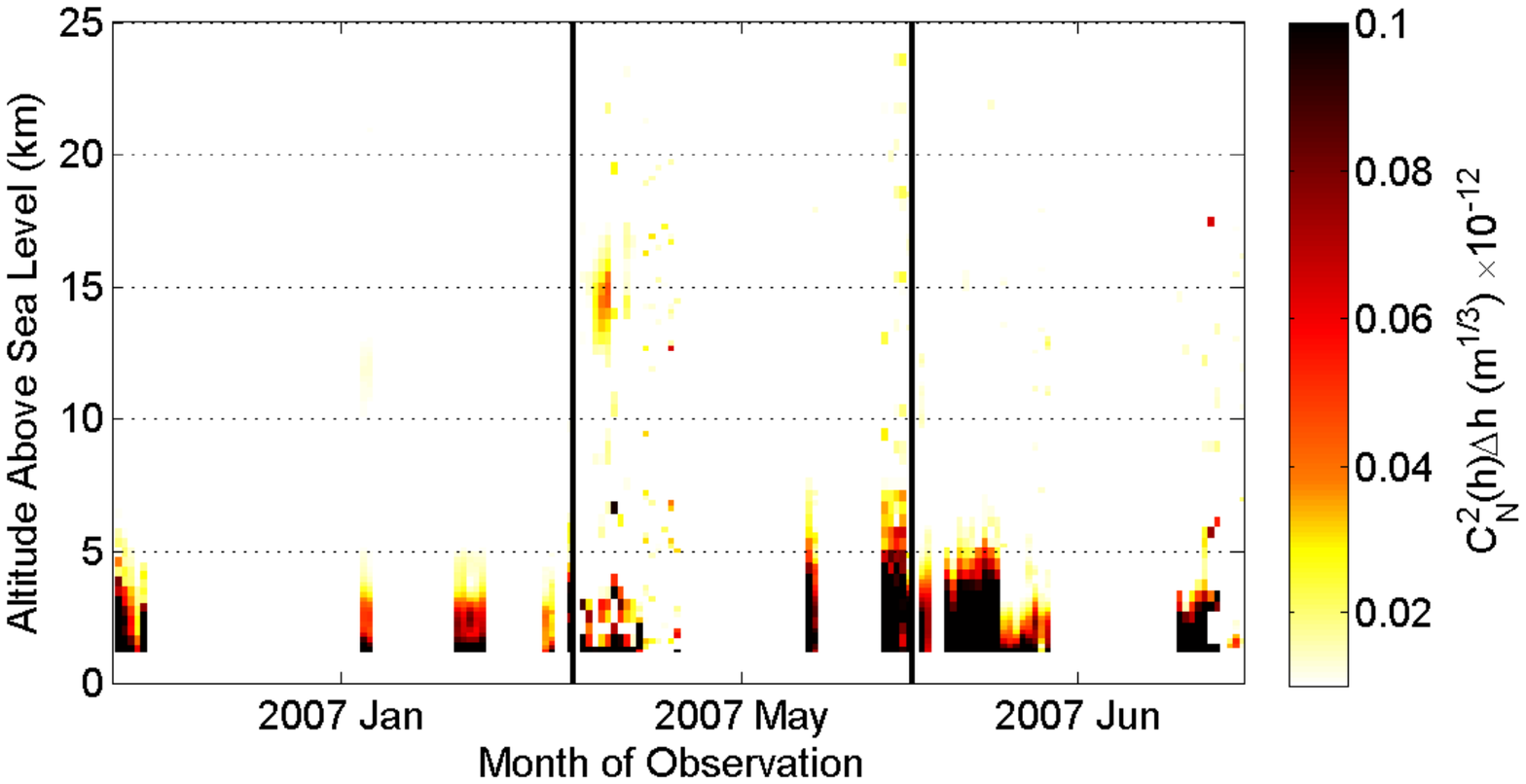}\label{fig:trending:2007_cn2G:scaled}}
  \subfigure[Scaled Corrected \Gen]{\includegraphics[width=\linewidth, trim=0mm 0mm 0mm 190mm]{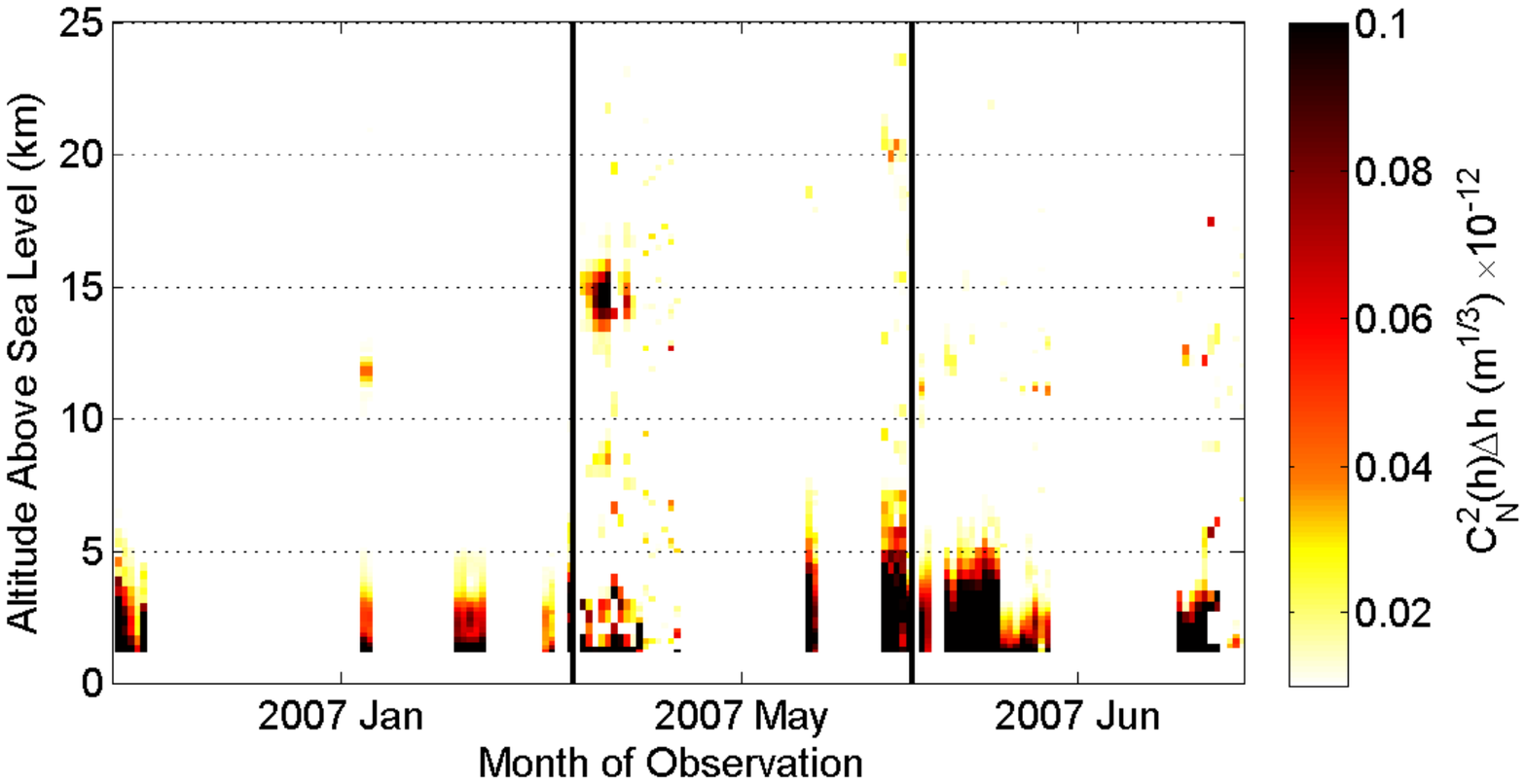}\label{fig:trending:2007_cn2G:scaledCorrected}} \caption[\Gen\s \cn\s profile trends observed over 2007.]{\Gen\s \cn\s profile trends observed over 2007. Time and \cn\s colour scaling used is as per Figure \ref{fig:trending:2007_cn2P}.  \subref{fig:trending:2007_cn2G:scaledCorrected} Correction factors applied to turbulent layers detected in the free atmosphere is incorporated.  See text for details.}\label{fig:trending:2007_cn2G}
\end{figure}

Figures \ref{fig:trending:2007_cn2P} and \ref{fig:trending:2007_cn2G} show the \pup\s and \gen\s \cn\s profiles respectively for data collected in 2007.  The NGT measurements at MJUO tend to dominate and typically mask any activity present in the upper layers, as seen in Figures \ref{fig:trending:2007_cn2P:norm} and \ref{fig:trending:2007_cn2G:norm}.  To reveal possible features in these upper layers the \cn\s profiles are scaled so that the colour range is limited to \cn\dh\s values between $10^{-14}$ and $10^{-13}$ \cndhunits.  Values below $10^{-14}$ \cndhunits\s are likely to result in images that are diffraction-limited rather than turbulence-limited on a 1-m telescope. $r_0$ is 1.55 m for a layer with $\cn\dh = 10^{-14}$ \cndhunits\s and a wavelength of 589 nm.  Values greater than $10^{-13}$ \cndhunits\s are approaching levels that are classified as strong turbulence \citep{AndrewsB47}.  Figure \ref{fig:trending:2007_cn2P:scaled}  shows the colour scaled images for \pup\s measurements from 2007.  Figure \ref{fig:trending:2007_cn2G:scaled} shows the colour scaled profiles for 2007 \gen\s measurements.

Examination of the overall \pup\s \cn\s measurements (Figure \ref{fig:trending:2007_cn2P}) indicates that there was significant turbulence located at low altitudes (i.e. below 5 km above sea level) with an additional layer that could be seen in some measurements at 12 -- 14 km above sea level.  Where low altitude turbulence was strong (i.e. all of January, the latter half of May and the majority of June) the upper altitude layers were masked.  In the \gen\s data (Figure \ref{fig:trending:2007_cn2G:scaled}) little to no upper altitude activity was detected.  The only exception occurred in the first half of May where little NGT was detected in the \pup\s data.  However the strength of the detected layer at approximately 14 km above sea level does not match the strength of the layer detected at comparable height in pupil-plane data.

From equations (\ref{eqn:tempT:covariance_binary}) and (\ref{eqn:secondaryCn2}), there is an assumption that each layer of turbulence present above a site is statistically independent and hence the strength of the layers found is not dependent on other layers in the structure.  If this assumption holds true then the measured strength of the detected turbulence from any given high altitude layer should be the same regardless of whether pupil-plane or generalised measurements were employed.  However issues arise under conditions of medium to strong NGT resulting in an underestimate of the strength of turbulence located in the higher levels in \gen\s SCIDAR measurements \citep{MohrPhD2009}.  A correction factor can be applied to the high altitude layers detected in \gen\s data.

Figure \ref{fig:trending:2007_cn2G:scaledCorrected} shows the colour scaled corrected \gen\s data for 2007.  The correction factor used was a localised Gaussian curve centred around the altitudes of the layers detected in the \pup\s data, correcting for differences in measurement plane, with a standard deviation of $2\dh$, where \dh\s is the altitude resolution of the measurement.   The level of correction needed is dependent on the altitude of the given layer and the strength of the NGT detected, based on simulations \citep{MohrPhD2009}.  It should be noted that this correction has little effect on the estimated $r_0$ for the profiles, due to the strength of the NGT layer that dominates the profiles.  Following the correction it can be noted that similar strength turbulent layers are seen in both the pupil-plane and generalised SCIDAR data for the high altitudes (Figures \ref{fig:trending:2007_cn2P:scaled} and \ref{fig:trending:2007_cn2G:scaledCorrected}). Where possible a similar correction has been applied to all generalised data presented in this paper.

The layer found at 12 -- 14 km above sea level can be associated with turbulence found in the tropopause region, which is commonly incorporated into models for \cn\s profiles \citep{HardyBook1998}.  For a site such as MJUO, a significant level of turbulence at low altitudes is expected.  MJUO is located roughly 50 km west of the Southern Alps.  With prevailing westerlies over much of New Zealand, the low- to mid- altitude wind structure is significantly affected by the terrain associated with the Southern Alps \citep{SturmanB49}.

\begin{figure}
  \centering
  \subfigure[Scaled \pup]{\includegraphics[width=\linewidth, trim=0mm 0mm 0mm 190mm]{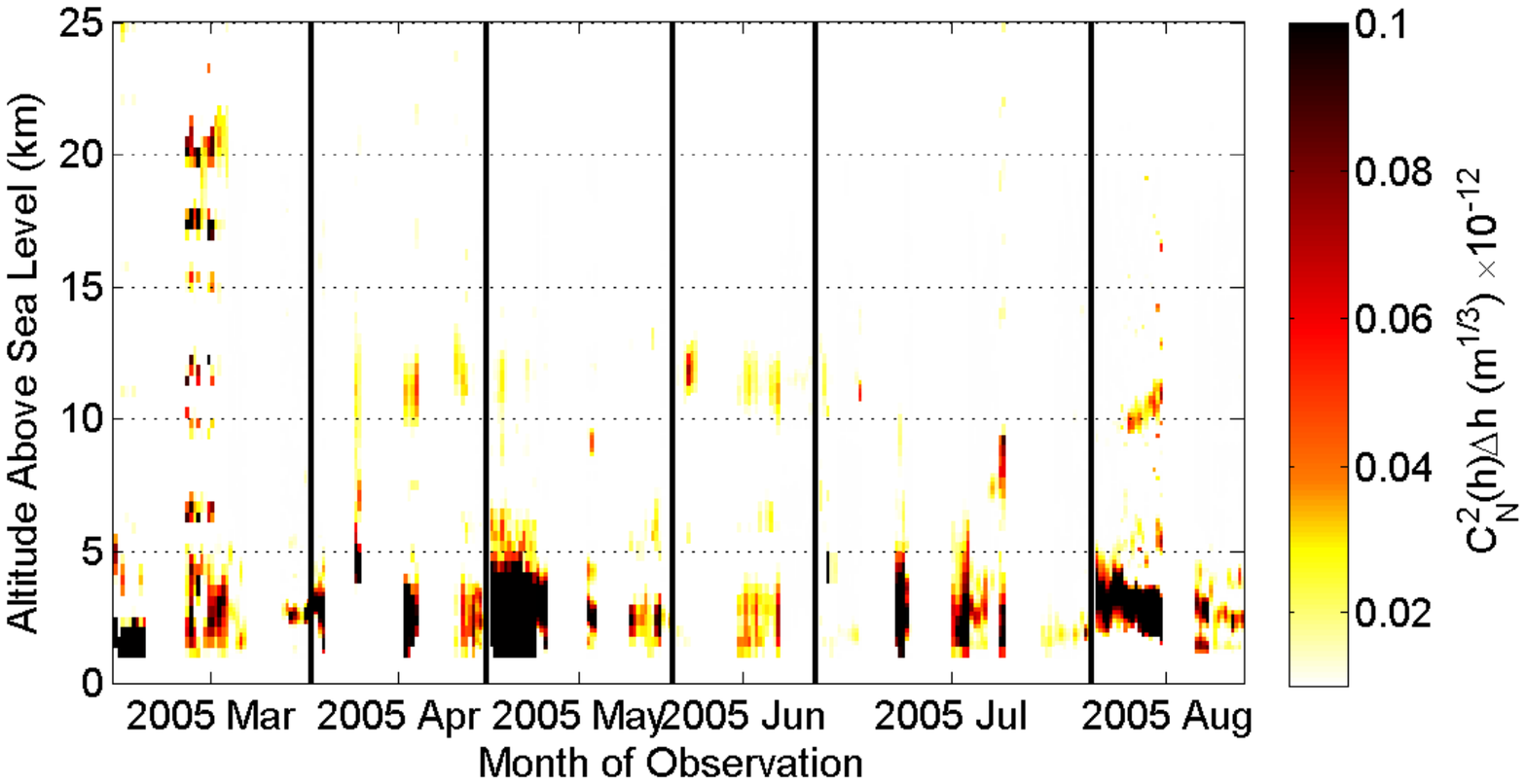}\label{fig:trending:2005_cn2P:scaled}}
  \subfigure[Scaled Corrected \Gen]{\includegraphics[width=\linewidth, trim=0mm 0mm 0mm 190mm]{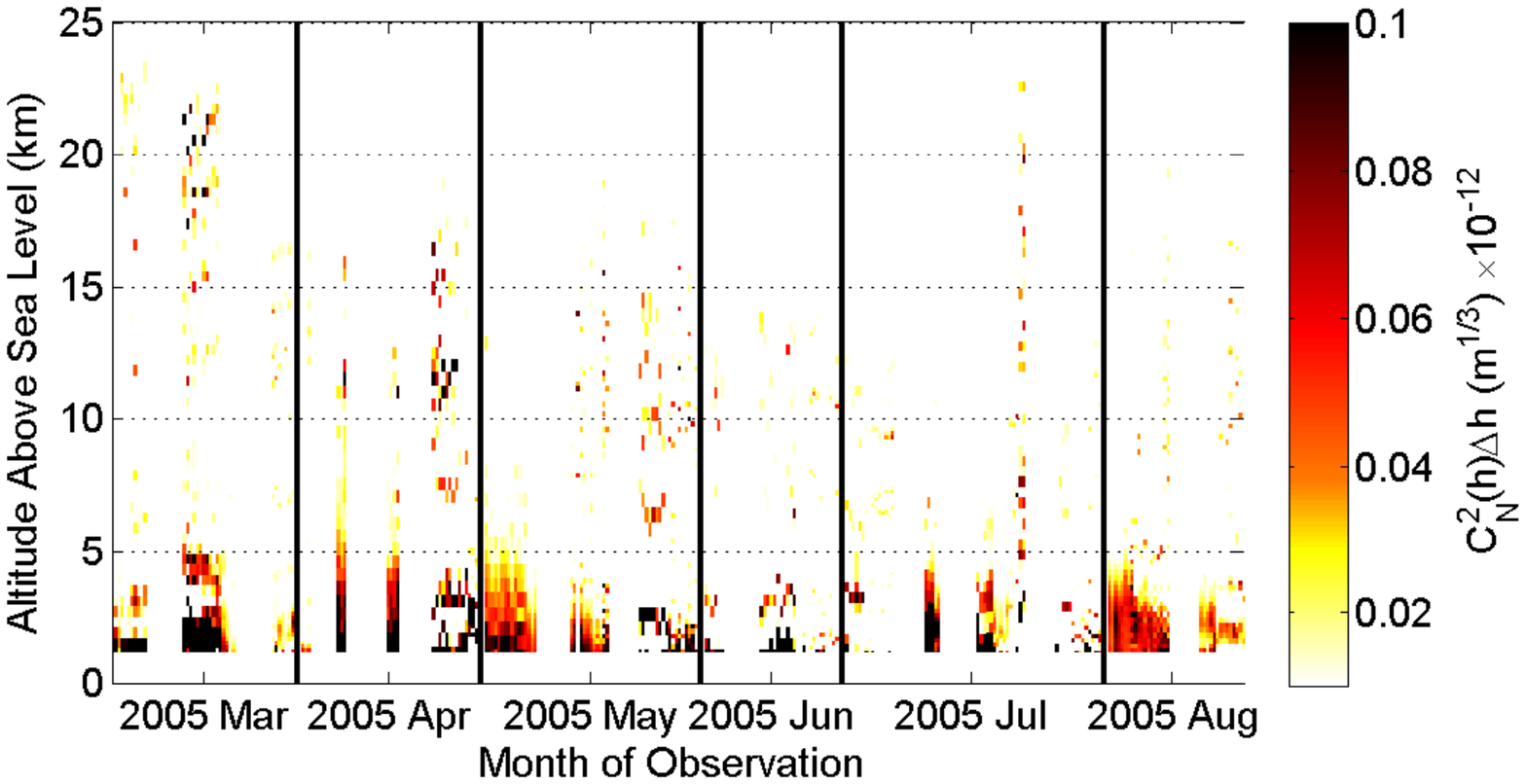}\label{fig:trending:2005_cn2G:scaled}}
  \caption{\Pup\s and \gen\s \cn\s profile trends observed over 2005. Time and \cn\s scaling, and \gen\s data corrections used are as per Figure \ref{fig:trending:2007_cn2G}.}\label{fig:trending:2005_cn2P}
\end{figure}

For data collected in 2005, shown in Figure \ref{fig:trending:2005_cn2P}, a strong low altitude layer located at less than 5 km above sea level is visible.  Also present is a weaker high altitude layer that ranges between 10 -- 14 km above sea level.  In some months an additional layer can be seen at 6 -- 8 km above sea level.

\begin{figure}[t]
  \centering
  \subfigure[Scaled \pup]{\includegraphics[width=\linewidth, trim=0mm 0mm 0mm 190mm]{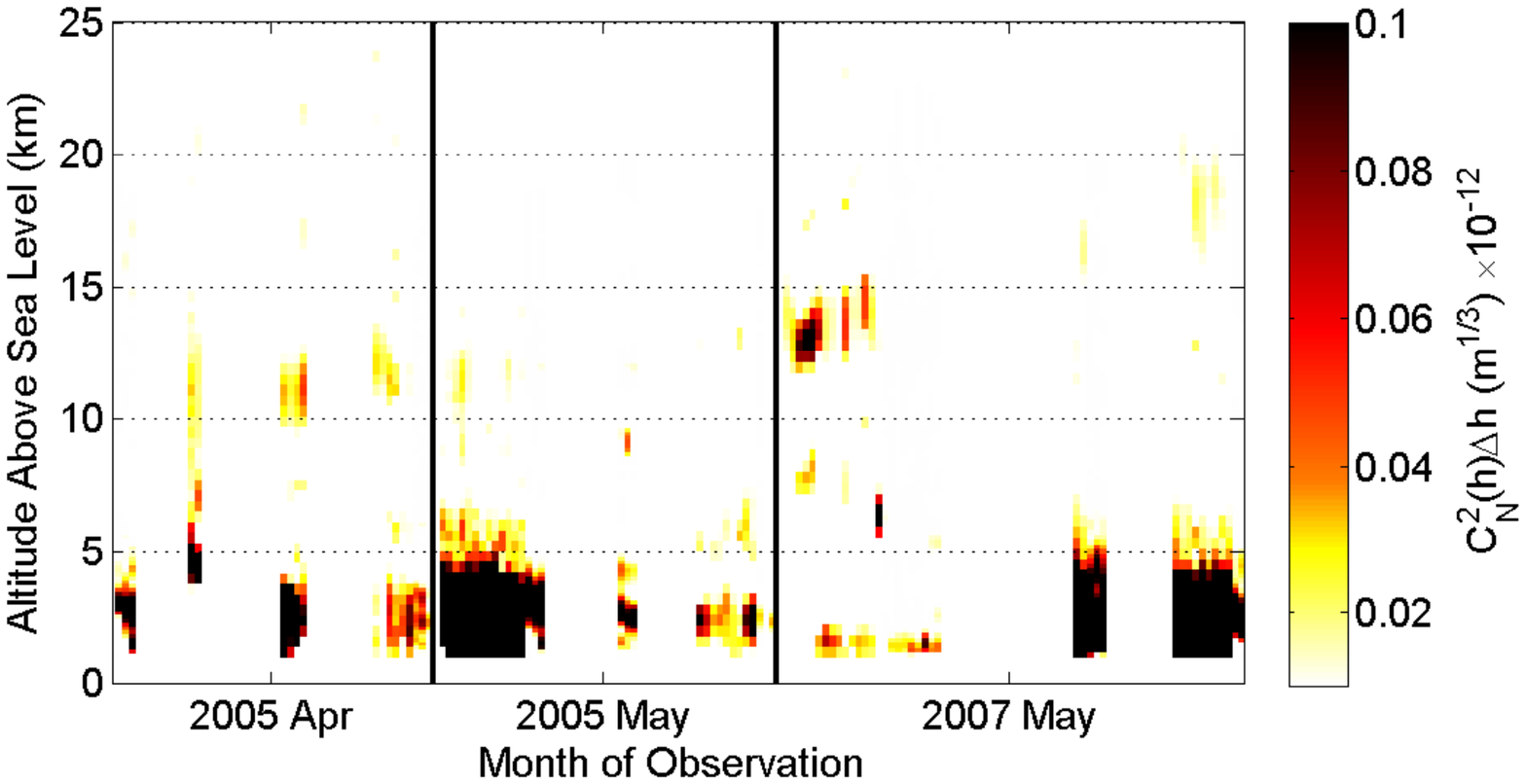}\label{fig:trending:fall2005through2007_cn2P:scaled}}
  \subfigure[Scaled corrected \gen]{\includegraphics[width=\linewidth, trim=0mm 0mm 0mm 190mm]{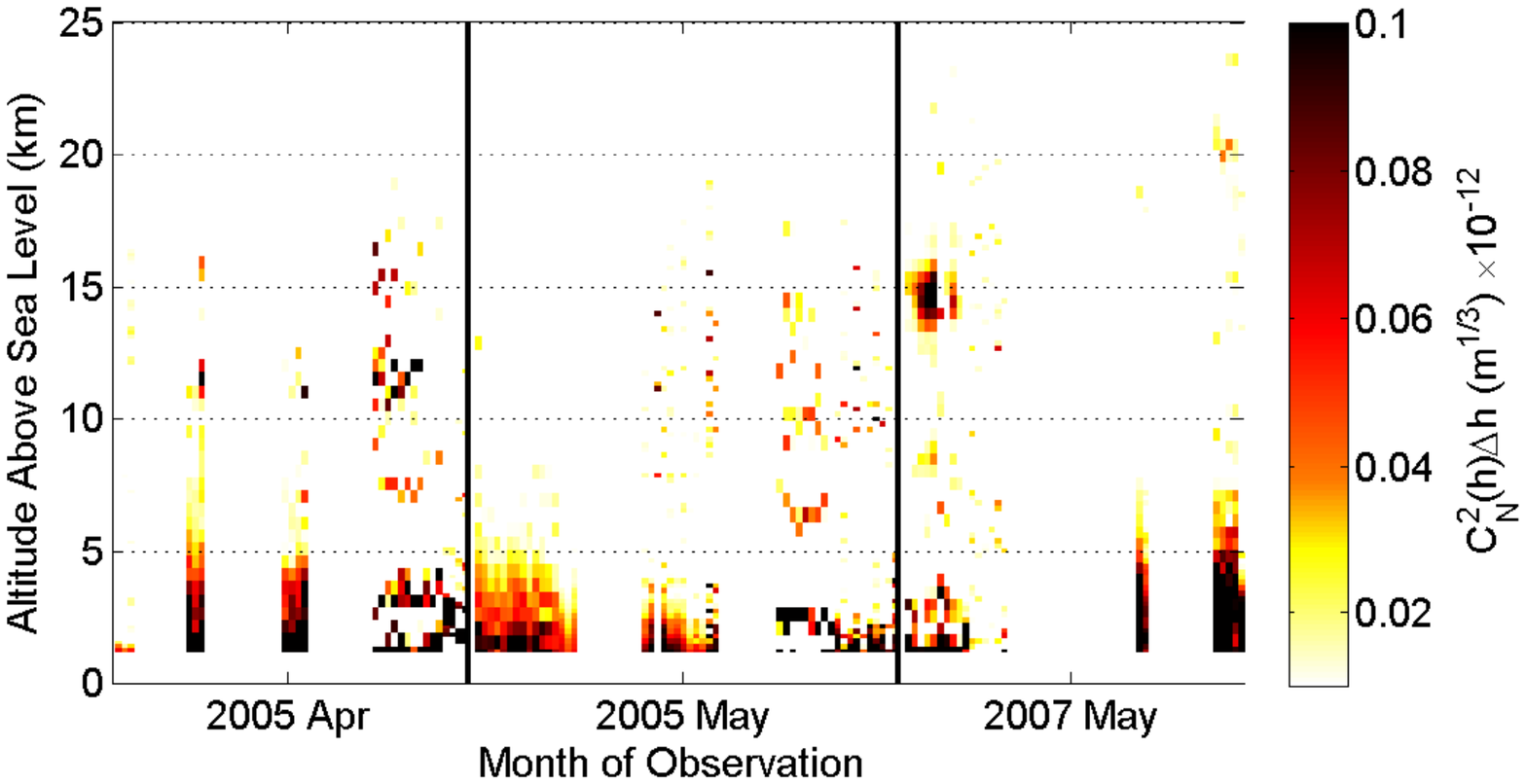}\label{fig:trending:fall2005through2007_cn2G:scaled}}
  \caption{\Pup\s and \gen\s \cn\s profile trends observed during autumn from 2005 to 2007. Time and \cn\s scaling, and \gen\s data corrections used are as per Figure \ref{fig:trending:2007_cn2G}.}\label{fig:trending:fall2005through2007_cn2P}
\end{figure}

Figure \ref{fig:trending:fall2005through2007_cn2P} shows the scaled \pup\s and \gen\s \cn\s profiles obtained for the autumn months (i.e. April and May) for UC-SCIDAR data collected in both 2005 and 2007.  During 2005 April and May, a weak high altitude layer was found at approximately 11 -- 12 km above sea level with a \cn\dh\s strength of approximately $3\times 10 ^{-14}$ \cndhunits.  In 2007 May the height of this high altitude layer increased by approximately 2 km.  The height of the tropopause is known to fluctuate up to 4 km throughout the year \citep{SturmanB49}.  It is reasonable to assume that this is turbulence generated at the same pressure scale height.  NGT and boundary layer turbulence is seen to extend up to 5 km above sea level.  This is seen in both the \pup\s and \gen\s data.  Little can be ascertained from the \gen\s data as to a trend in the height and strength of the high altitude layer, due to the strength of the NGT and the noise present in the 2005 data.

\begin{figure}
  \centering
  \subfigure[Scaled \pup]{\includegraphics[width=\linewidth, trim=0mm 0mm 0mm 190mm]{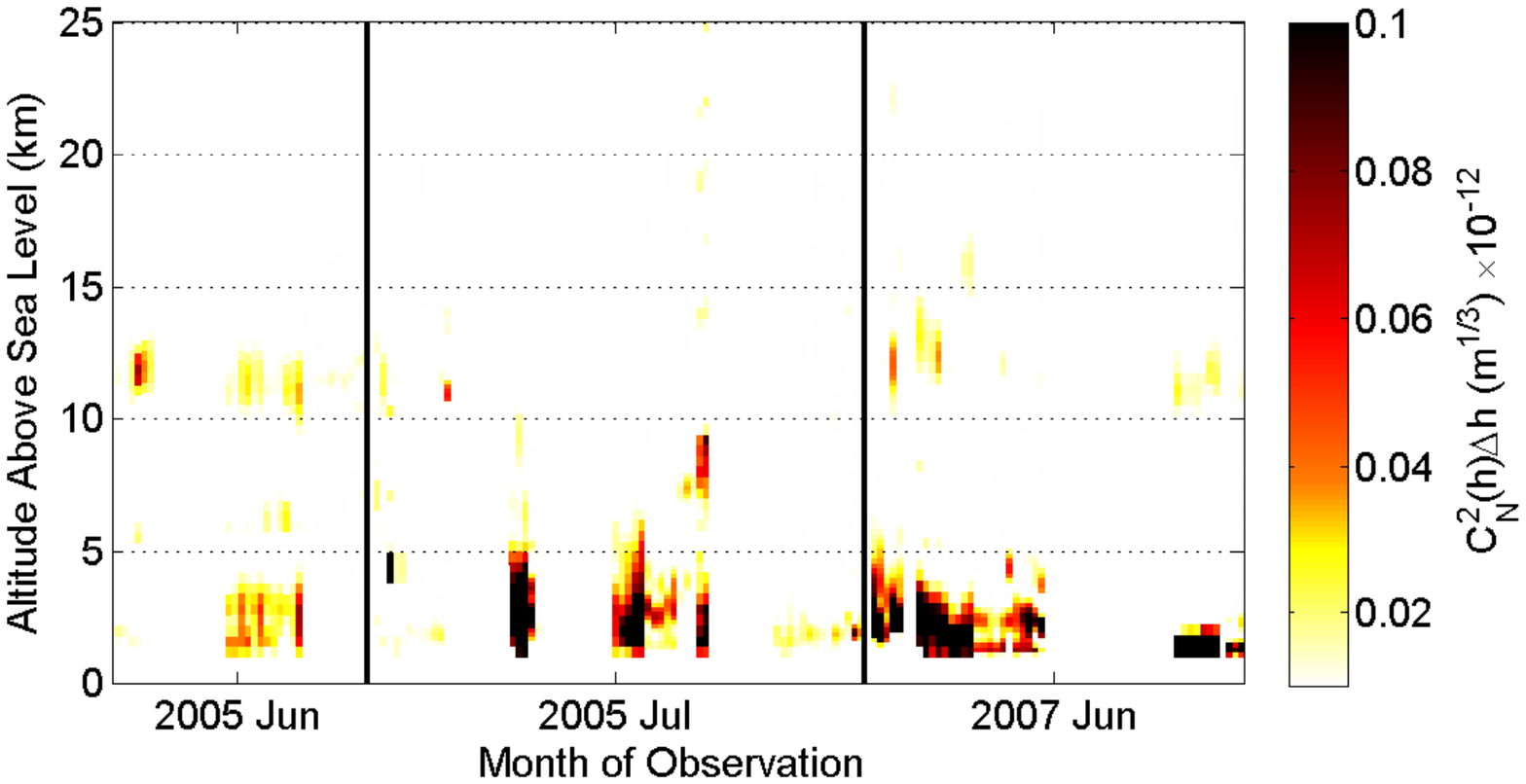}\label{fig:trending:winter2005through2007_cn2P:scaled}}
  \subfigure[Scaled Corrected \gen]{\includegraphics[width=\linewidth, trim=0mm 0mm 0mm 190mm]{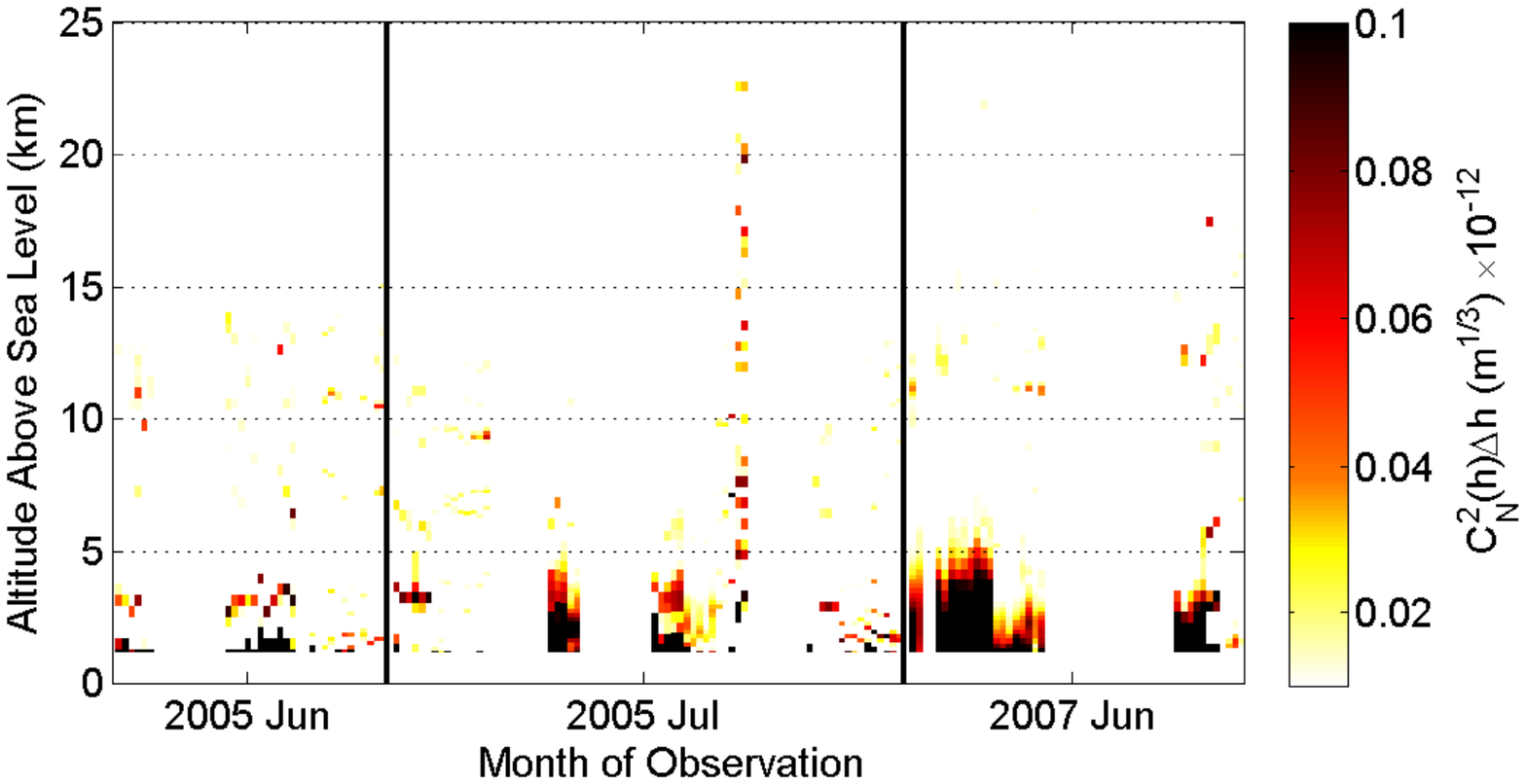}\label{fig:trending:winter2005through2007_cn2G:scaled}}
  \caption{\Pup\s and \gen\s \cn\s profile trends observed during winter from 2005 to 2007. Time and \cn\s scaling, and \gen\s data corrections used are as per Figure \ref{fig:trending:2007_cn2G}.}\label{fig:trending:winter2005through2007_cn2P}
\end{figure}

Figure \ref{fig:trending:winter2005through2007_cn2P} shows the scaled \cn\s profiles obtained for the winter months (i.e. June and July) with UC-SCIDAR collected in both 2005 and 2007.  A high altitude layer is seen consistently at approximately 12 km above sea level with an average \cn\dh\s strength of approximately $3\times 10^{-14}$ \cndhunits.  This is of similar strength to that seen in the autumn months (i.e. April and May), suggesting little to no difference in the high altitude layer.  However, an additional mid-altitude layer is seen in some data, with heights ranging from 6 -- 8 km above sea level  and with varying strength.  As seen with the April/May data, low altitude turbulence extends up to 5 km above sea level.  Again little can be ascertained from the \gen\s data about trends in the heights and strengths of the high altitude layers.

The monthly averages and standard deviations for $r_0$ and $\theta_0$ are summarised in Table \ref{tab:trending:all_r0theta0}.  Also shown is the mean turbulence altitude, $\overline{h_0}$, which can be found using \citep{GarciaLorenzoMNRAS2009}
\begin{equation}\label{eqn:h0}
    \overline{h_0} = 0.314\frac{r_0}{\theta_0},
\end{equation}
providing a measure of the effective height of dominant turbulence if the multi-layer structure was replaced with a single layer.  Note that $\overline{h_0}$ values shown reflect the distances above sea level.  For generalised data, $\overline{h_0}$ has been corrected for defocus distances for each observation sequence.

January 2007 data exhibited similar $r_0$ and $\theta_0$ values for both \pup\s and \gen\s data, which indicates a significant low-altitude layer.  This is also shown in the $\overline{h_0}$ values of 4.8 and 2.9 km for pupil-plane and generalised data respectively.  In summer, the longer, warmer days heat the surrounding ground and buildings which can lead to greater NGT effects.  The \pup\s $\theta_0$ values obtained for May are smaller than those obtained for June although the $r_0$ values are similar for the two months.  The difference in $\theta_0$ can be attributed to the high altitude layers found in May, which were stronger and higher than those found in June (Figure \ref{fig:trending:2007_cn2G:scaled}).  This is also seen in the higher pupil-plane $\overline{h_0}$ value for 2007 May of 7.3 km compared to 5.7 km for 2007 June.   

During 2005, the winter months (i.e. June and July) have a \pup\s \averzero\s in the order of 20 cm, whereas autumn (i.e. March, April and May) and spring (i.e. August) have a \pup\s \averzero\s ranging from approximately 10 cm to 15 cm.  \avethetazero\s follows a similar trend to that of \averzero, with the winter months having a value of approximately 2 arcsec compared to 1.3 -- 1.7 arcsec for autumn and spring.  \avethetazero\s for 2005 March was 1.6 arcsec with a large standard deviation \stdthetazero\s of 0.7.  This is associated with the noise present in a significant portion of the runs during 2005 March.  The \averzero\s values for \gen\s measurements are reasonably consistent throughout the two years; 5 -- 6 cm for 2007, and 7 cm for 2005.  Little variation is seen for \gen\s \avethetazero\s values across the entire campaign, suggesting a consistent, dominating NGT layer.  The significant variations seen in \pup\s \averzero\s and \avethetazero\s values suggest that the strengths of the high altitude layers fluctuate.  The average $r_0$ for all UC-SCIDAR data was $12\pm5$ cm and $7\pm1$ cm for the \pup\s and \gen\s measurements respectively.  $\theta_0$ was $1.5\pm0.5$ arcsec and $1.0\pm0.1$ arcsec for the \pup\s and \gen\s measurements respectively.  $\overline{h_0}$ was $6\pm1$ km and $2.0\pm0.7$ km for pupil-plane and generalised measurements respectively.  The values of $\overline{h_0}$ obtained does reinforce that the dominating turbulence at MJUO is located at near ground altitudes.

UC-SCIDAR estimates for $r_0$ are smaller than those obtained in 1999 April using the Imperial College system (i.e. $12.3\pm1.4$ cm) \citep{JohnstonJ44}.  Although layer height estimates were similar for the two systems, the strength of the NGT layer was in the order of 6 -- 10 times stronger for UC-SCIDAR data.  Some variation can be expected due to the amount of time that has passed between the two instruments being used.  However the lower $r_0$ values obtained using UC-SCIDAR do match the observation conditions typically seen by observers, which has a nominal angular resolution, $\theta_{\mathrm{res}}$, of $\sim2$ arcsec (A. Gilmore (MJUO) 2006, private communication).  $\theta_{\mathrm{res}}$ from UC-SCIDAR was 2.5 arcsec for the full profile, calculated at a wavelength of 589 nm. $\theta_{\mathrm{res}}$ for data taken in 1999 April was 1.2 arcsec.

The large variation in the \pup\s measurements suggests not only a possible relationship with seasonal changes, but also with the weather at the site.  June and July of 2005 saw high \pup\s\averzero\s and \avethetazero\s values which can be attributed to the calmer weather seen during these observational periods.  However most other months show similar \averzero\s and \avethetazero\s estimates within the margin of error.  This suggests that weather conditions have a greater influence on the profiles obtained.  It is suggested that data obtained from UC-SCIDAR be correlated to meteorological data to investigate whether SCIDAR data could predict weather-related seeing over the site.

\begin{table*}
  \centering
  \caption{Monthly averages and standard deviations for $r_0$, $\theta_0$ and $h_0$ for all months. $\overline{h_0}$ values are indicated as distances above the sea level.  Elevation of MJUO is 1024 m. Generalised values have been computed based on corrected profiles.  Generalised values of $\overline{h_0}$ have been corrected for defocus distances.}\label{tab:trending:all_r0theta0}
\begin{center}
\begin{tabular}{c|ccccc|ccccc}
\hline
  &  \multicolumn{4}{c}{\Pup } &   &  \multicolumn{4}{c}{\Gen } &  \\
 Month &  $\overline{r_0}$ &  $\sigma_{r_0}$ &  $\overline{\theta_0}$ &  $\sigma_{\theta_0}$ &  $\overline{h_0}$ &  $\overline{r_0}$ &  $\sigma_{r_0}$ &  $\overline{\theta_0}$ &  $\sigma_{\theta_0}$ & $\overline{h_0}$ \\
  &  (cm) &  (cm) &  (arcsec) &  (arcsec) &  (km) &  (cm) &  (cm) &  (arcsec) &  (arcsec) & (km) \\
\hline
 2007 June &  12 &  3 &  1.8 &  0.7 &  5.7 &  5 &  0 &  1.1 &  0.3 & 2.3 \\
 2007 May &  11 &  7 &  1.3 &  0.9 &  7.3 &  6 &  1 &  1.0 &  0.2 & 2.6 \\
 2007 January &  6 &  3 &  0.8 &  0.2 &  4.8 &  5 &  2 &  0.9 &  0.1 & 2.9 \\
\hline
 2005 August &  10 &  2 &  1.7 &  0.4 &  4.5 &  8 &  1 &  1.1 &  0.1 & 1.6 \\
 2005 July &  18 &  4 &  2.2 &  0.4 &  6.3 &  7 &  1 &  1.2 &  0.2 & 1.3 \\
 2005 June &  22 &  3 &  1.8 &  0.3 &  8.8 &  6 &  1 &  1.0 &  0.1 & 1.5 \\
 2005 May &  12 &  1 &  1.7 &  0.2 &  5.0 &  7 &  1 &  0.9 &  0.2 & 1.1 \\
 2005 April &  11 &  3 &  1.3 &  0.3 &  6.8 &  7 &  1 &  0.9 &  0.1 & 1.9 \\
 2005 March &  15 &  3 &  1.6 &  0.7 &  7.2 &  8 &  2 &  1.1 &  0.3 & 2.8 \\
\hline
 1999 April &  -- &  -- &  -- &  -- & --  &  12.3 &  1.4 &  -- &  -- & -- \\
\hline
\end{tabular}\end{center}
\end{table*}

\subsection{\vw\s Profiles}

\begin{figure}
  \centering
  \includegraphics[width=\linewidth, trim=0mm 0mm 0mm 190mm]{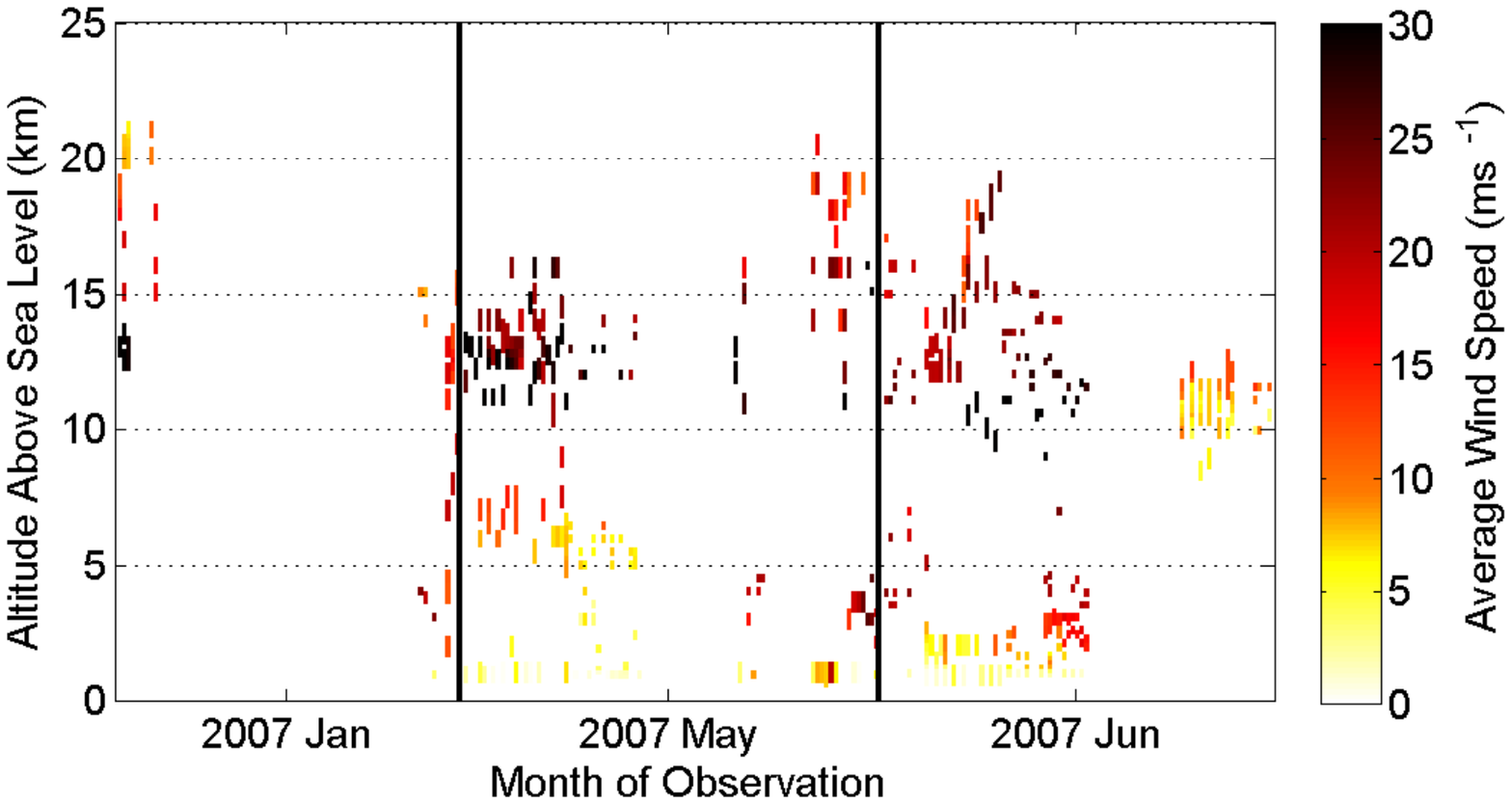}
  \caption[Average wind speeds, \vwmag, for observations taken during 2007.]{Average wind speeds, \vwmag, for observations taken during 2007.  Measurements from the individual months are separated by solid black lines.}\label{fig:trending:2007_vh}
\end{figure}

Figure \ref{fig:trending:2007_vh} shows the average wind speed, \vwmag, measured as a function of height, $h$, for the 2007 data.  There is some scatter, due to errors associated with height estimation, but the measurements show a turbulent layer at 11 -- 14 km above sea level with an average speed of 18 \vhunits, with velocities ranging from $\sim6.5$ \vhunits\s to over 30 \vhunits.  In some cases a mid-altitude layer at 6 -- 7 km above sea level can be seen moving at approximately 7 \vhunits.  On nights with significant NGT, a low-altitude layer was seen with wind speeds ranging between 10 to 24 \vhunits.

Velocity measurements obtained in April 1999 indicated high-altitude layers at 11 and 13 km travelling at 6.45 and 11.63 \vhunits\s respectively.  This is consistent with the measurements obtained using UC-SCIDAR for layers at this altitude.

Table \ref{tab:trending:fg} shows the monthly averages and standard deviations for \fg.  Note that the value shown for  2007 January is averaged across both the \pup\s and \gen\s data.  The values for 2007 June reflect the average for data collected on May 31 and June 1 only.

\begin{table}[h]
  \centering
  \caption{Nightly averages for \fg\s for all months.  Generalised values have been computed based on corrected profiles.}\label{tab:trending:fg}
  \begin{tabular}{c|cc|cc}
  \hline
  &  \multicolumn{2}{c}{\Pup } &  \multicolumn{2}{c}{\Gen } \\
  Month &  $\overline{f_G}$  &  $\sigma_{f_G}$ &  $\overline{f_G}$  & $\sigma_{f_G}$  \\
    & (Hz) & (Hz) & (Hz) & (Hz) \\
  \hline
2007 June  &  26	&	12	&	29	&	31	\\
2007 May &  30	&	23	&	8	&	13	\\
2007 January  & 20 &  15 &  -- & -- \\
 \hline
2005 July  & 10	&	7	&	11	&	14	\\
2005 June  & 13	&	5	&	10	&	18	\\
  \hline
\end{tabular}
\end{table}

The monthly calculated Greenwood frequency averages, $\overline{f_G}$, for 2007 were approximately 30 Hz or less.  For a wavelength of 589 nm, the estimate for \fg\s ranges between 30 -- 90 Hz depending on the models used for \cn\s and \vw\s profiles \citep{TysonB48}.  The measured \fg\s obtained at MJUO is at the bottom end of this range.  It is most likely that the \fg\s has been underestimated due to the gaps that exist in the \vh\s profiles.

\fg\s calculated from SCIDAR can be underestimated because wind velocity measurements are reliant on covariance strengths in the measurement plane being sufficiently strong with respect to the background covariance noise.  Aperture normalisation in the data amplifies noise, especially near the aperture edge.  Peaks approaching the aperture edge can be hidden by the noise.  The detection of only the central peak with one secondary peak, termed partial triplet analysis \citep{MohrIVCNZ08}, does allow for more layer velocities to be found.  In addition, some covariance peaks may be obscured due to their close proximity to other peaks. Although all layers seen in \vw\s will have an associated \cn\s strength,  not all \cn\s layers will have a measurable \vw\s due to the position of the covariance peaks relative to other peaks and the aperture edge and the resulting covariance strength.  This results in gaps in the measured \vw\s profile and is reflected in the large $\sigma_{f_G}$ values.

The UC-SCIDAR system used in 2005 captured data at a frame rate of 30 Hz, which limits the maximum detectable velocity on a 1-m telescope to 30 \vhunits\s under ideal conditions \citep{MohrSPIE2008}.  This was a limitation of the CCD cameras used at the time.  Based on the \vh\s profiles from the current system, which has a frame rate of 60 Hz, data collected during 2005 should provide a reasonable \vh\s profile for low- to mid-altitude layers, however high-altitude layer velocities may not be measurable.  Temporal analysis was performed on data from June and July 2005 data only.  These months provided the most reliable data set with a large number of runs utilising exposures of 1 -- 2 ms.

\begin{figure}
  \centering
  \includegraphics[width=\linewidth, trim=0mm 0mm 0mm 190mm]{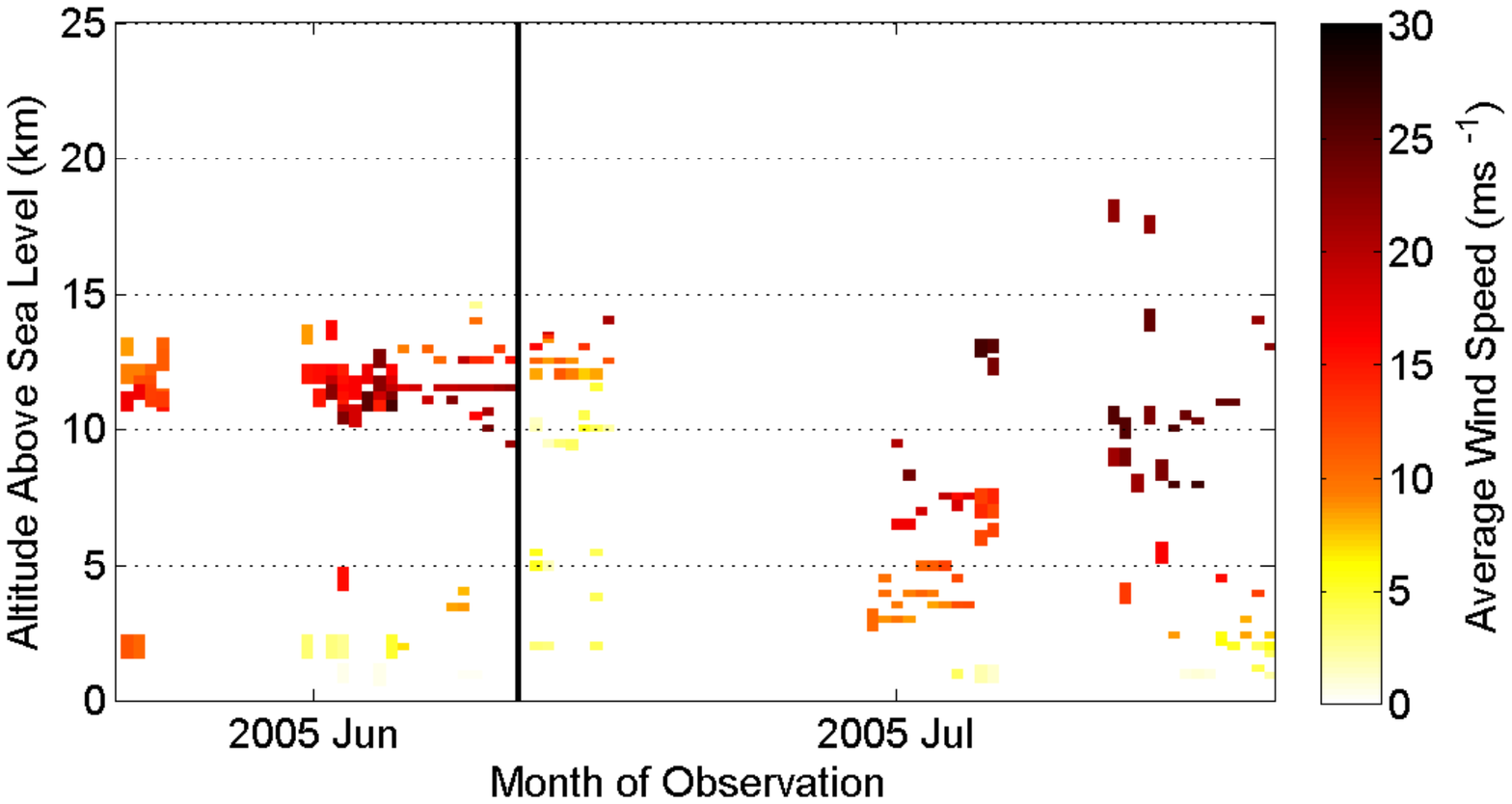}
  \caption{As per Figure \ref{fig:trending:2007_vh}, except for 2005 data.}\label{fig:trending:2005_vh}
\end{figure}

Figure \ref{fig:trending:2005_vh} shows the average wind speeds for 2005 June and July.  In June a layer at approximately 12 km above sea level was moving consistently at 12 -- 15 \vhunits.  In July there was much more scatter in the average velocity.

Using the \vh\s profiles obtained for 2005 June and July, the average Greenwood frequency, \avefg, was found to range between 10 and 20 Hz.  There are significant gaps present in the \vh\s profiles, as indicated by the standard deviation in \fg, \stdfg, ranging between 5 and 15 Hz (Table \ref{tab:trending:fg}).  As such \avefg\s is likely to be underestimated.

\begin{figure}
  \centering
  \includegraphics[width=\linewidth, trim=0mm 0mm 0mm 190mm]{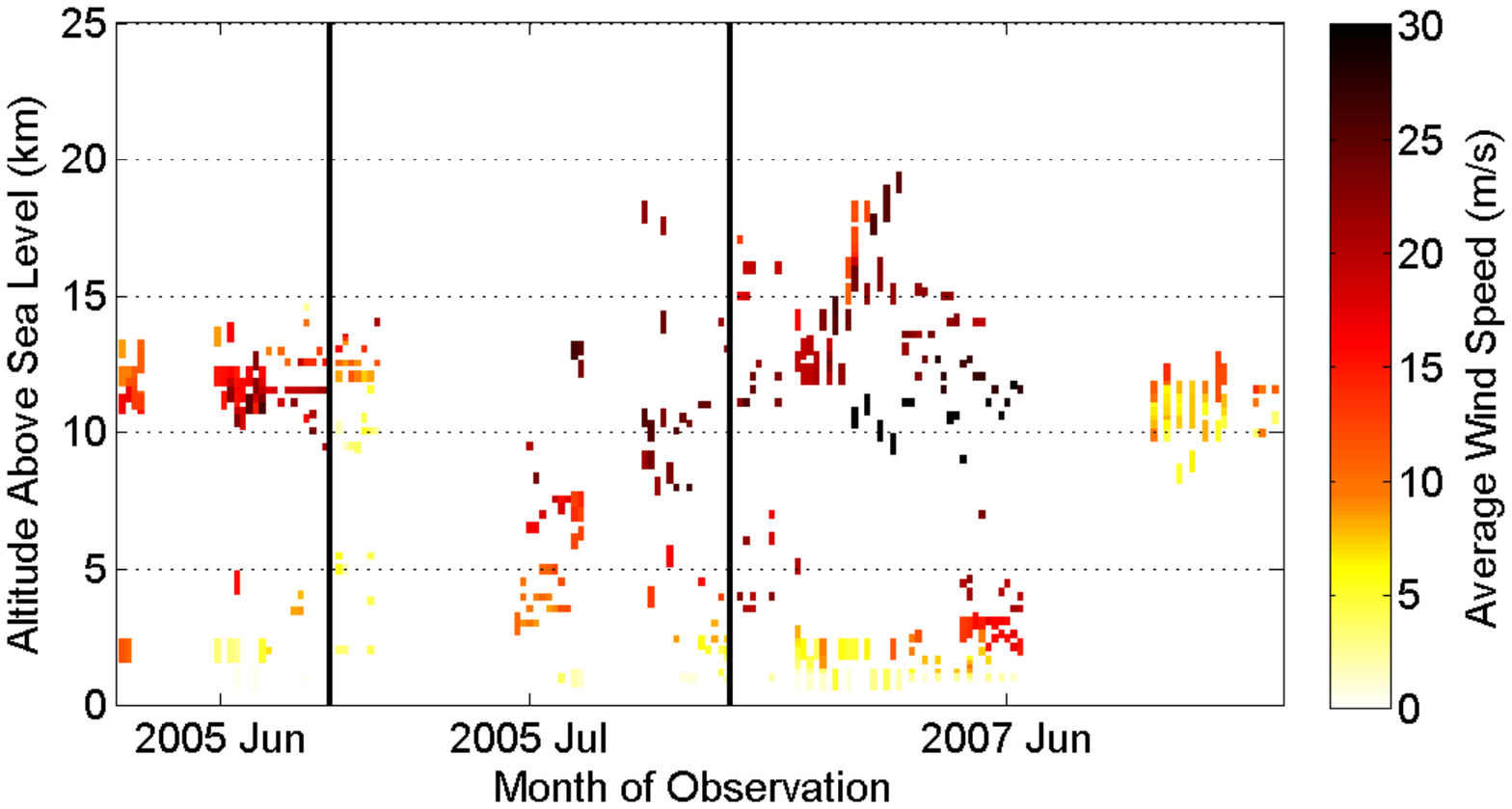}
  \caption{As per Figure \ref{fig:trending:2007_vh}, except for observations taken during winter months of 2005 and 2007.}\label{fig:trending:winter2005through2007_vh}
\end{figure}

Figure \ref{fig:trending:winter2005through2007_vh} shows the average wind speed for the winter months (June and July).  Although a significant amount of scatter exists for the data from 2005 July and 2007 June the layer heights found are similar.  The high altitude layer has an average speed of 10 -- 15 \vhunits\s in calmer weather, but speeds in excess of 25 \vhunits\s at other times. 


\section{Finding a Model that Fits}
\label{sec:model}
\subsection{\cn\s Model}
Optical turbulence is highly irregular, where turbulence strengths can vary by an order of magnitude around a mean value \citep{HardyBook1998}.  Commonly used models represent a mean profile of \cn\s measurements taken over extended periods of time.

The Hufnagel-Valley (HV) model is commonly used to describe the average turbulence at an astronomical site \citep{TysonB48}.  A standard HV model consists of three main components: an exponentially decreasing \cn\s through the troposphere; a peak at approximately 10 km above ground corresponding to a tropo-pause layer; and a strong surface layer \citep{HardyBook1998}.

At many sites additional layers have been detected at low- to mid-troposphere altitudes \citep{AvilaJ267, FuensalidaC89, PrieurJ58, WangJ330}.  The generic HV model, incorporating an additional mid-altitude layer, is a sum of exponential terms such that \cn\s is given by \citep{HardyBook1998}
\begin{eqnarray}\label{eqn:trending:HVmodel}
    \cn & =&  A\exp\left(-\frac{h}{H_A}\right)\nonumber \\
    && + B\exp\left(-\frac{h}{H_B}\right)\nonumber\\
      & & + C h^{10}\exp\left(-\frac{h}{H_C}\right) \nonumber \\
      &&+ D\exp\left(-\frac{(h - H_D)^2}{2d^2}\right),
\end{eqnarray}
where $A$ is the turbulence coefficient for near-ground turbulence (i.e. $\propto C_N^2(0)$) and $H_A$ is the height for its $1/e$ decay, $B$ and $H_B$ are similarly defined for turbulence in the troposphere, and $C$ and $H_C$ are related to the turbulence peak located at the tropopause.  The fourth term in equation (\ref{eqn:trending:HVmodel}) can be used to define one or more isolated layers, where $D$ and $H_D$ define the strength and height of the layer and $d$ specifies the layer thickness.

\begin{figure}
\centering
  \subfigure[Selected \pup\s and \gen\s data]{\includegraphics[width=0.8\linewidth, trim=0mm 70mm 0mm 90mm]{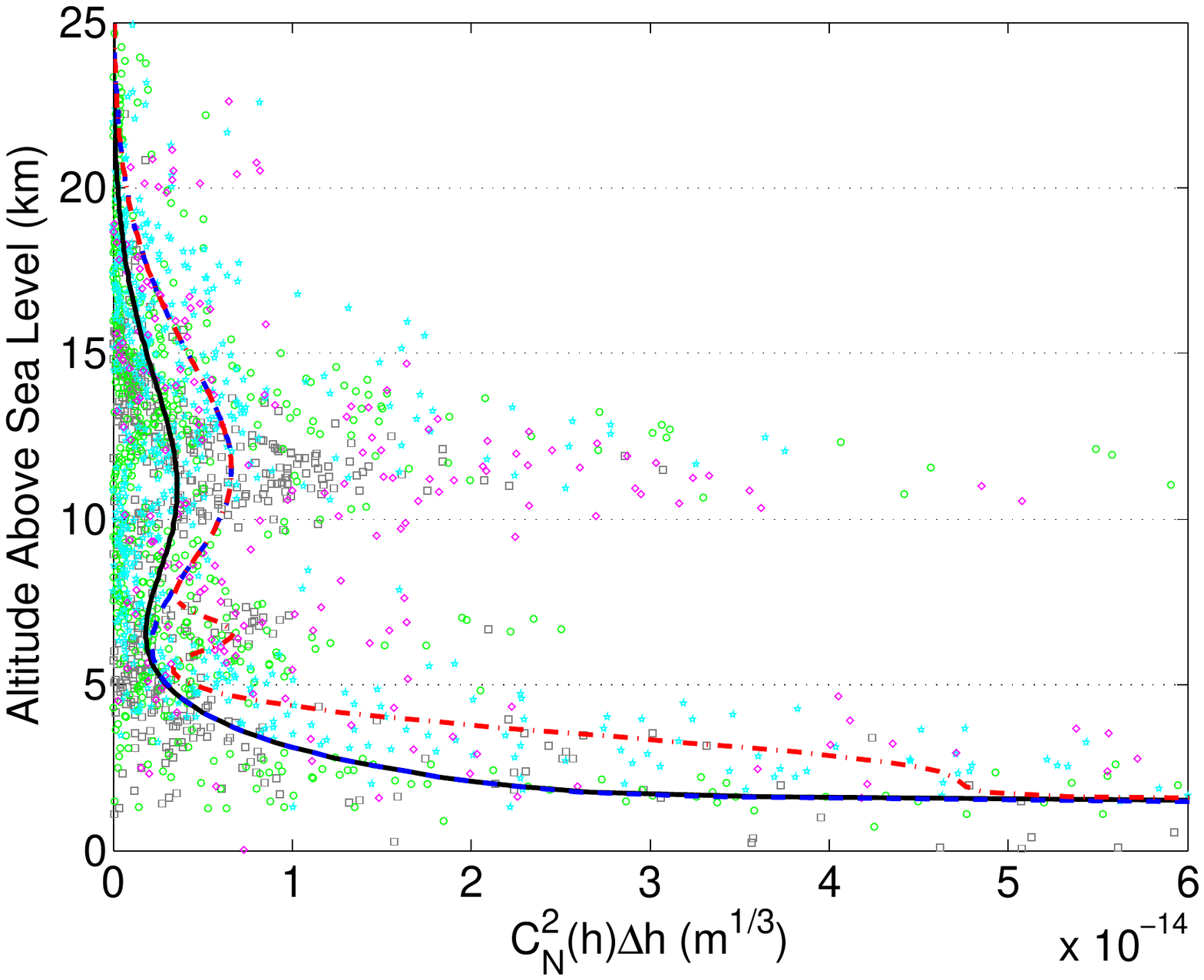}\label{fig:trending:modelfit:p}}\\
  \subfigure[Semi-logarithmic plot of \subref{fig:trending:modelfit:p}]{\includegraphics[width=0.8\linewidth, trim=0mm 70mm 0mm 90mm]{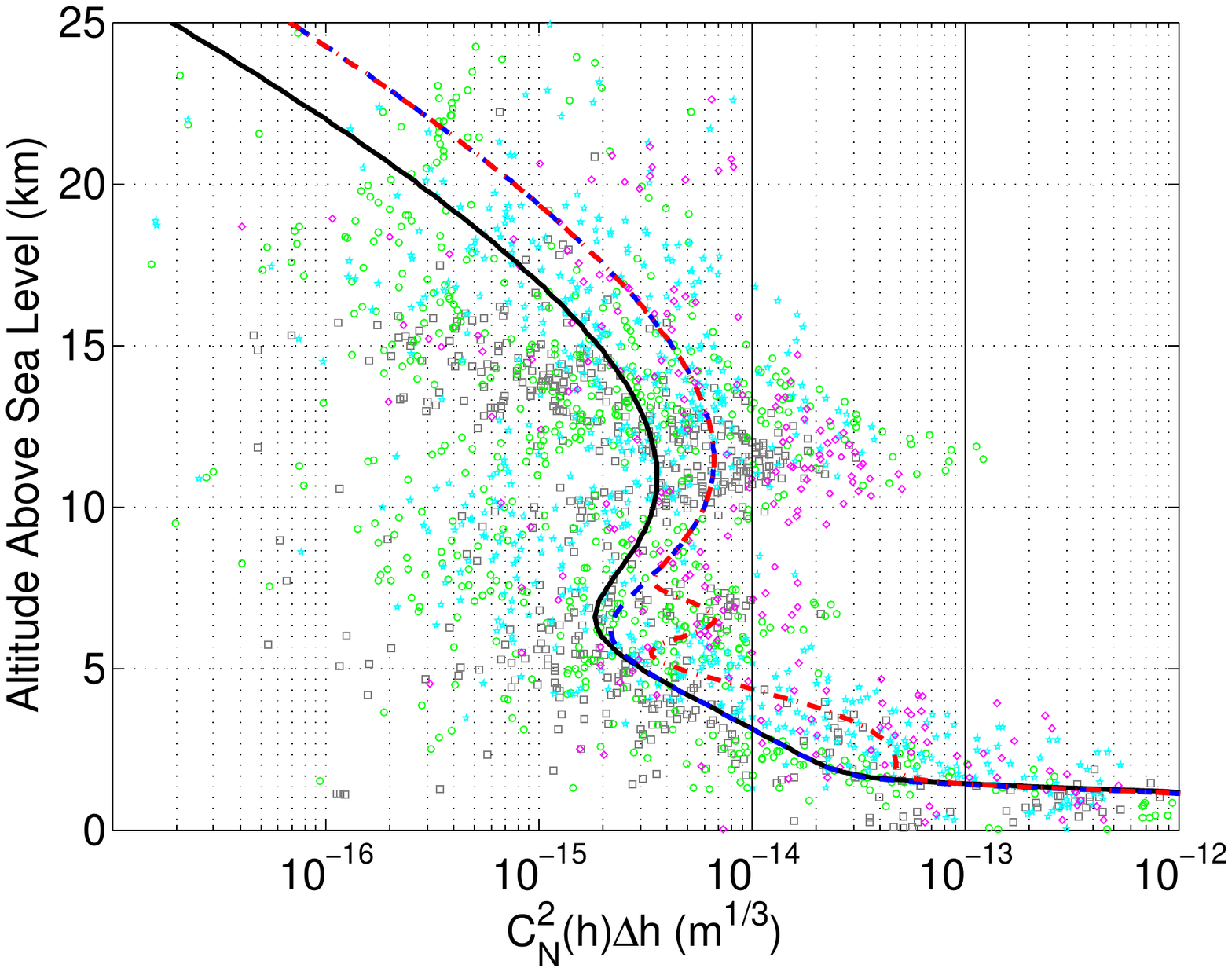}\label{fig:trending:modelfit:p_log}}\\
  \caption[\cn\s model fitting.]{\cn\s model fitting.  Both \subref{fig:trending:modelfit:p} and \subref{fig:trending:modelfit:p_log} show selected \pup\s and \gen\s data from (\textcolor{cyan}{$\bigstar$}) 2007 June, (\textcolor{green}{$\bigcirc$}) 2007 May, (\textcolor{grey}{$\square$}) 2005 June and (\textcolor{magenta}{$\lozenge$}) 2005 April.  Models shown are (---) the HV 5-7 model, (\textcolor{blue}{-- --}) MJUO1: a modified HV model and (\textcolor{red}{-- $\cdot$ --}) MJUO3: a modified HV model that incorporates two additional layers.}\label{fig:trending:modelfit}
\end{figure}

Figure \ref{fig:trending:modelfit} shows selected \pup\s and \gen\s \cn\s profiles obtained from various months in 2005 and 2007.  Also shown are three different models all based on the HV model.  Table \ref{tab:trending:cn2modelParams} lists the parameters for the various models shown in Figure \ref{fig:trending:modelfit}.  The HV 5-7 model (indicated by the solid black line) has parameters such that the resulting $r_0$ and $\theta_0$ are 5 cm and 7 $\mu$radians (1.44 arcsec) when using $\lambda = 500$ nm.  The HV 5-7 model, while producing $r_0$ and $\theta_0$ values appropriate for the site (i.e. 6 cm and 1.7 arcsec respectively for $\lambda = 589$ nm),  results in a tropopause layer that is slightly too low, residing at 10 km above the site (i.e. 11 km above sea level) and too weak.  By increasing the altitude of the tropopause peak by 500 m and increasing the coefficient $C$ from $3.59 \times 10^{-53}$ to $5.94 \times 10^{-53}$ (MJUO1 model) then the $\theta_0$ decreases to 0.96 arcseconds which is more in line with the measured values from the \gen\s \cn\s profiles.

\begin{figure}
  \centering
  \includegraphics[width=0.8\linewidth, trim=0mm 70mm 0mm 90mm]{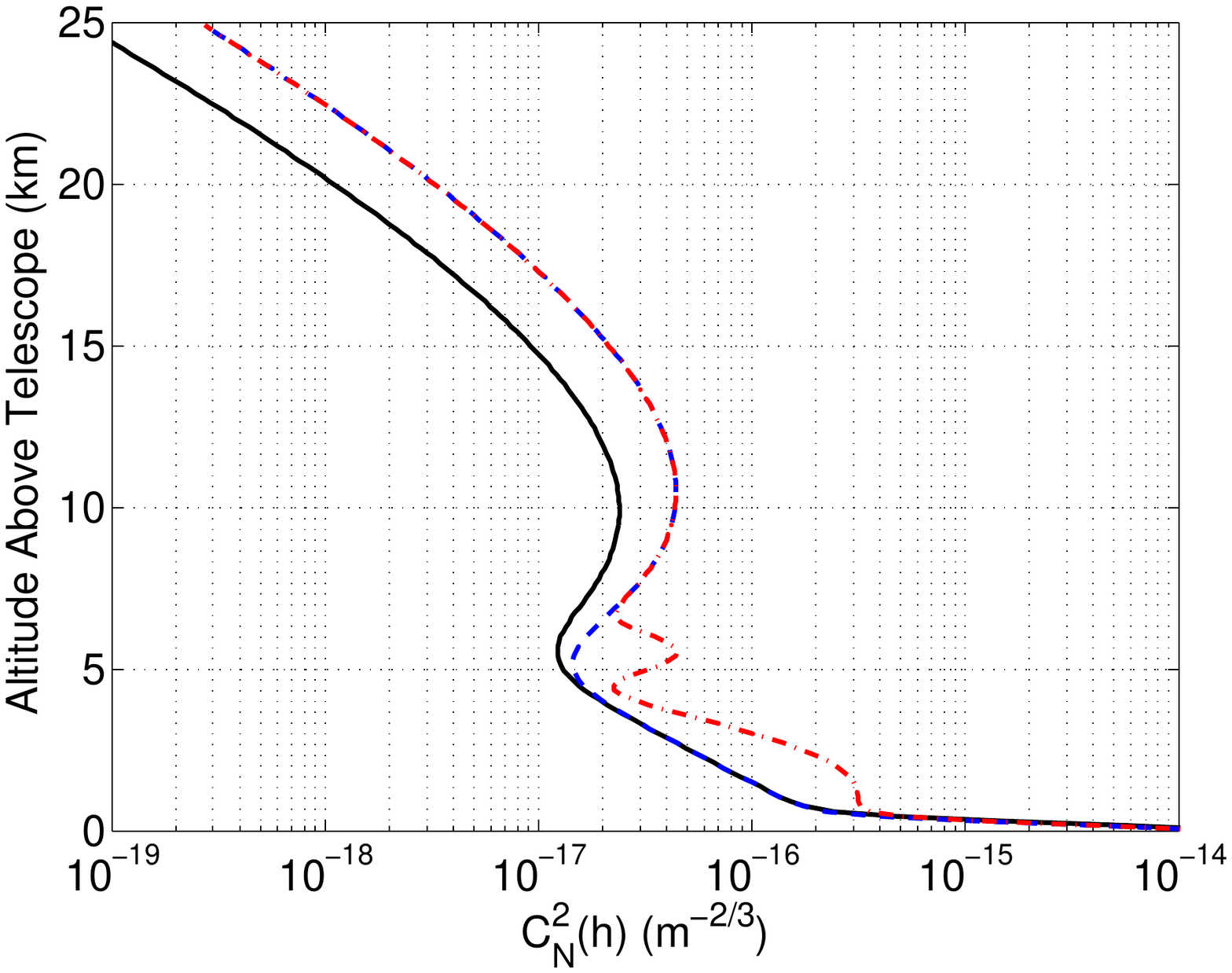}
  \caption[\cn\s turbulence models for MJUO.]{\cn\s turbulence models for MJUO. (---) HV 5-7 model; (\textcolor{blue}{-- --}) MJUO1: Modified HV model; (\textcolor{red}{-- $\cdot$ --}) MJUO3: Modified HV model incorporating two additional layers.  MJUO3 is the recommended model.}\label{fig:trending:Cn2Model}
\end{figure}

The near-ground and low-altitude turbulence is seen to regularly extend up to 5 km above sea level (Figures \ref{fig:trending:2007_cn2P} -- \ref{fig:trending:winter2005through2007_cn2P}).  An isolated layer was added with a peak at 1.5 km above the site (i.e. 2.5 km above sea level) and a thickness of 1 km.  The peak strength of this layer has been set to $2\times 10^{-16}$ \cnunits.  The modified HV model, with the addition of this low-altitude layer, is called MJUO2.  In some profiles an additional layer was found at heights ranging from 6 -- 8 km above sea level.  MJUO3 incorporates an additional layer at 6.5 km above sea level, and is the recommended model for use in the AO design for MJUO.  The profiles of the HV 5-7 model, MJUO1 and MJUO3 are shown in Figure \ref{fig:trending:Cn2Model}.

Additional layers at approximately 2 and 6.5 km about sea level have also be regularly detected at other sites.  \citet{AvilaJ65} reported the detection of five different layers during the 1998 site testing campaign of Cerro Tololo and Cerro Pach\'{o}n in Chile.  Strong layers were found at the tropopause region (i.e. 11 -- 12 km) and at low altitudes, extending up to approximately 4 km above sea level.  Although the layer found at 6.5 km was often weaker than the low altitude turbulence, it was consistently present.  Similar low- to mid- altitude turbulence was detected at San Pedro M\'{a}rtir, Mexico \citep{AvilaJ326}.

The HV model is such that the resulting peak \linebreak present at the tropopause region is not a spike but rather a broad peak, which accounts for variations in layer heights and fluctuations in the turbulence strength seen over time. Instantaneous spikes and variations in the heights of the turbulent layers are smoothed out.  This has the effect of broadening the peaks seen, particularly in the tropopause region.  
Although it would be ideal to refine the model such that the breath of the tropopause region is not so wide, as a model for AO design MJUO3 is the recommended model.  Using a wavelength of 589 nm, $r_0$ is estimated to be 6 cm for MJUO3.  $\theta_0$ is estimated to be 0.9 arcseconds.  
$\overline{h_0}$ is estimated at 4.2 km above the measurement plane.  If a defocus distance of 3 km was used the resulting $\overline{h_0}$ of 1.2 km would be in line with the values obtained from the generalised UC-SCIDAR measurements.

\begin{table*}
\begin{minipage}[c]{\linewidth}
  \begin{center}
  \caption{Parameters for \cn\s turbulence models using a generic HV model.}\label{tab:trending:cn2modelParams}
\small{\begin{center}\begin{tabular}{cccccccc}
\hline
 Model\footnote{For all models indicated $A = 17\times 10^{-15}$, $H_A = 100$ m, $B = 27 \times 10^{-17}$, and $H_B = 1500$ m.} &  $C$ &  $H_C$ &  $D$ &  $H_{D}$ &  $d$ &  $r_0$\footnote{$r_0$ and $\theta_0$ values are specified for $\lambda = 589$ nm for the full profile and for $h > 3$ km above the telescope in brackets.} & $\theta_0$ \\
        &  ($\times 10^{-53}$) &  (m) &  ($\times 10^{-16}$) &  (m) &  (m) &  (cm) & (arcsec) \\
\hline
 HV 5-7 &  3.59 &  1000 &  0 &   &   &  6 (27) & 1.74 (1.81) \\
 MJUO1  &  5.94 &  1050 &  0 &   &   &  6 (16) & 0.94 (0.96) \\
 MJUO2\footnote{MJUO2 includes a strong low-altitude layer.}  &  5.94 &  1050 &  2 &  1500 &  1000 &  5 (16) & 0.92 (0.95) \\
 MJUO3\footnote{MJUO3 incorporates an additional mid-altitude layer for which the parameters are indicated in brackets.}  &  5.94 &  1050 &  2 (0.3) &  1500 (5500) &  1000 (500) &  5 (15) & 0.90 (0.94) \\
\hline
\end{tabular}\end{center}}
\end{center}
\end{minipage}
\end{table*}

\subsection{\vw\s Models}
To describe the wind velocity with increasing altitude, a Greenwood wind model is commonly used \citep{HardyBook1998}.  The Greenwood model is a Gaussian based model and is defined as \citep{TysonB48}
\begin{eqnarray}\label{eqn:trending:GreenwoodVelocity}
    \vw &= & V(0) + V(H_T)\exp\left[ -\left(\frac{h\cos\zeta - H_T}{L_T}\right)^2\right]\nonumber\\
    &&\times\left[\sin^2\beta +\cos^2\beta\cos^2\zeta\right]^{1/2},
\end{eqnarray}
where $V(0)$ is the wind velocity at ground level, $V(H_T)$ is the velocity at the tropopause located at an altitude $H_T$, $L_T$ is the thickness of the tropopause layer, $\beta$ is the wind direction relative to the telescope azimuth and $\zeta$ is the angle of the telescope from zenith.  The direction corrections used in the Greenwood wind model are for a telescope that uses a horizontal coordinate system and strictly speaking should not be applied to telescopes at MJUO as they employ an equatorial coordinate system.  However for AO design it is the wind speed that is important.  For the purpose of further discussion it will be assumed that $\beta = 0\degree$ and hence equation (\ref{eqn:trending:GreenwoodVelocity}) becomes
\begin{equation}\label{eqn:trending:GreenwoodVelocityBetaZero}
    \vw = V(0) + V(H_T)\exp\left[ -\left(\frac{h\cos\zeta - H_T}{L_T}\right)^2\right]\times \cos\zeta.
\end{equation}

\begin{figure}
  \centering
  \subfigure{\includegraphics[width=\linewidth, trim=0mm 60mm 30mm 150mm]{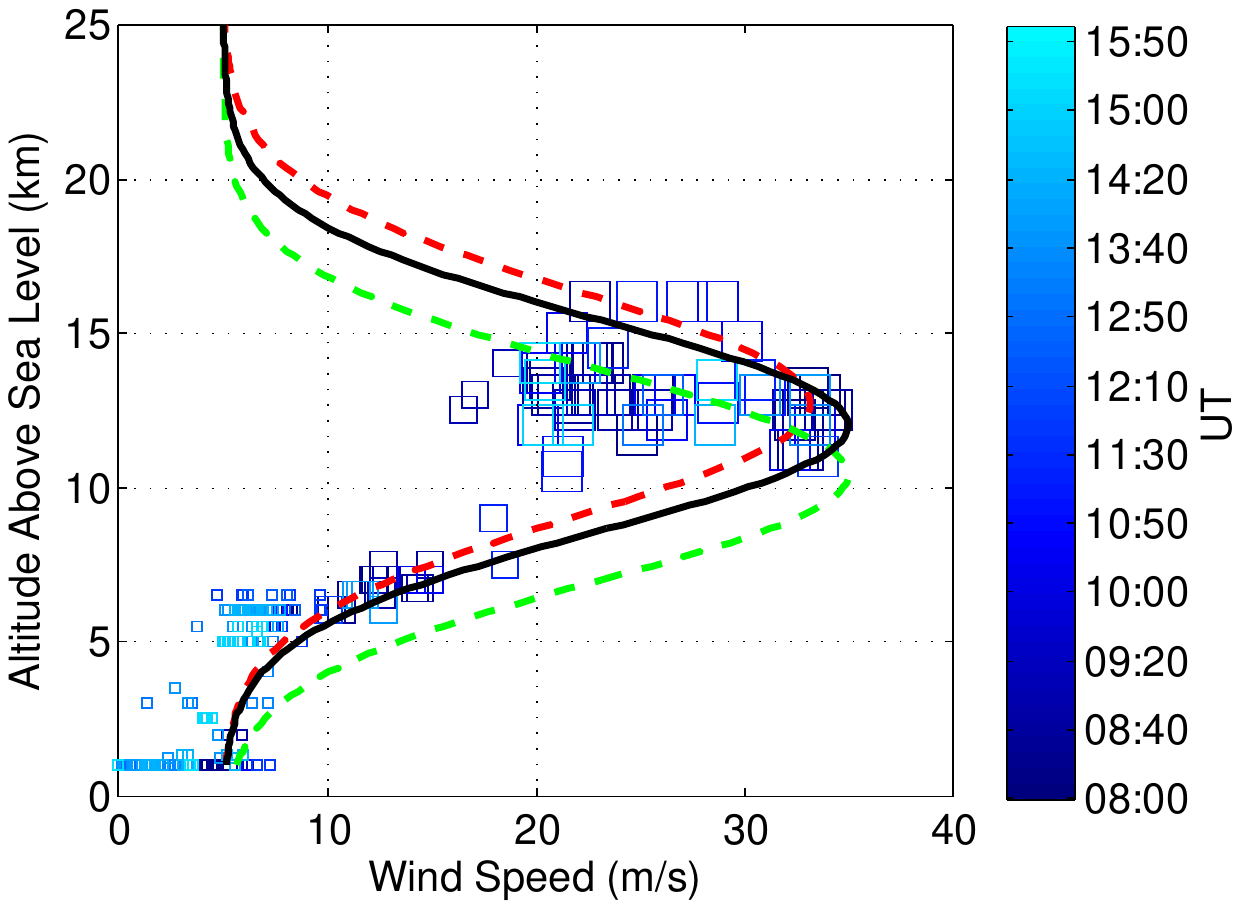}}
\subfigure{\includegraphics[width=\linewidth, trim=0mm 150mm 0mm 90mm]{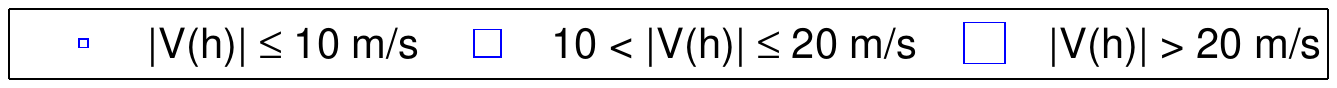}}
  \caption[Wind speed analysis for observations made on 2007 May 3.]{Wind speed analysis for observations made on 2007 May 3.  Models shown are (\textcolor{green}{- -}) the standard Bufton model, (---) a modified Bufton model with $H_T = 11$ km, and (\textcolor{red}{- -}) a modified Bufton model with $H_T = 11$ km and $\zeta = 20\degree$. Elevation for MJUO is 1024 m above sea level.}\label{fig:trending:may07_vel_model}
\end{figure}

Figure \ref{fig:trending:may07_vel_model} shows the \vw\s measurements from 2007 May 3.  The \vw\s profile for these data has an ideal profile that can be modelled by a modified Bufton wind model, where the Bufton model is a specific Greenwood model where $V(0) = 5$ \vhunits\s and $V(H_T) = 30$ \vhunits\s with $H_T = 9.4$ km and $L_T = 4.8$ km for $\zeta = 0\degree$.  Also shown in Figure \ref{fig:trending:may07_vel_model} is the Bufton and modified Bufton wind models.  The standard Bufton wind model (indicated by the dashed green line) assumes that the tropopause layer is at 9.4 km above the telescope with a wind speed of 30 \vhunits.  This is low for the \vw\s profile for May 3.  Moving the model tropopause height to 12 km above sea level (indicated by the solid black line) allows the model to encompass the lower turbulent layers detected, as well as the activity detected in the tropopause region.  If a zenith angle was incorporated into the model (red dashed line) then the effective wind speed in the tropopause region decreases to a value that is better suited to the data.  The modified Bufton model used incorporated a zenith angle of 20\degree, as this was the angle used by most of the measurements made on May 3.

Regardless of the model parameters used, a Gaussian based model will not adequately describe the \vw\s profiles detected when low altitude layers at 2 -- 5 km have wind velocities between 15 -- 20 \vhunits.  A large portion of the data collected for velocity had the presence of some low-altitude wind.  It was decided to add a second Gaussian peak, such that
\begin{eqnarray}\label{eqn:trending:ModifiedGreenwoodVelocity}
    \vw & = &  V(0) \\
        &   &  + V(H_T)\exp\left[ -\left(\frac{h\cos\zeta - H_T}{L_T}\right)^2\right]\times \cos\zeta \nonumber \\
        &   &  + V(H_1)\exp\left[ -\left(\frac{h\cos\zeta - H_1}{L_1}\right)^2\right]\times \cos\zeta,\nonumber
\end{eqnarray}
where $V(H_1)$ is the velocity of a low-altitude layer located at $H_1$ above the telescope with a thickness of $L_1$.

\begin{figure}
  \centering
  \subfigure[Fit of \vw\s models to UC-SCIDAR data]{\includegraphics[width=0.8\linewidth, trim=0mm 70mm 0mm 90mm]{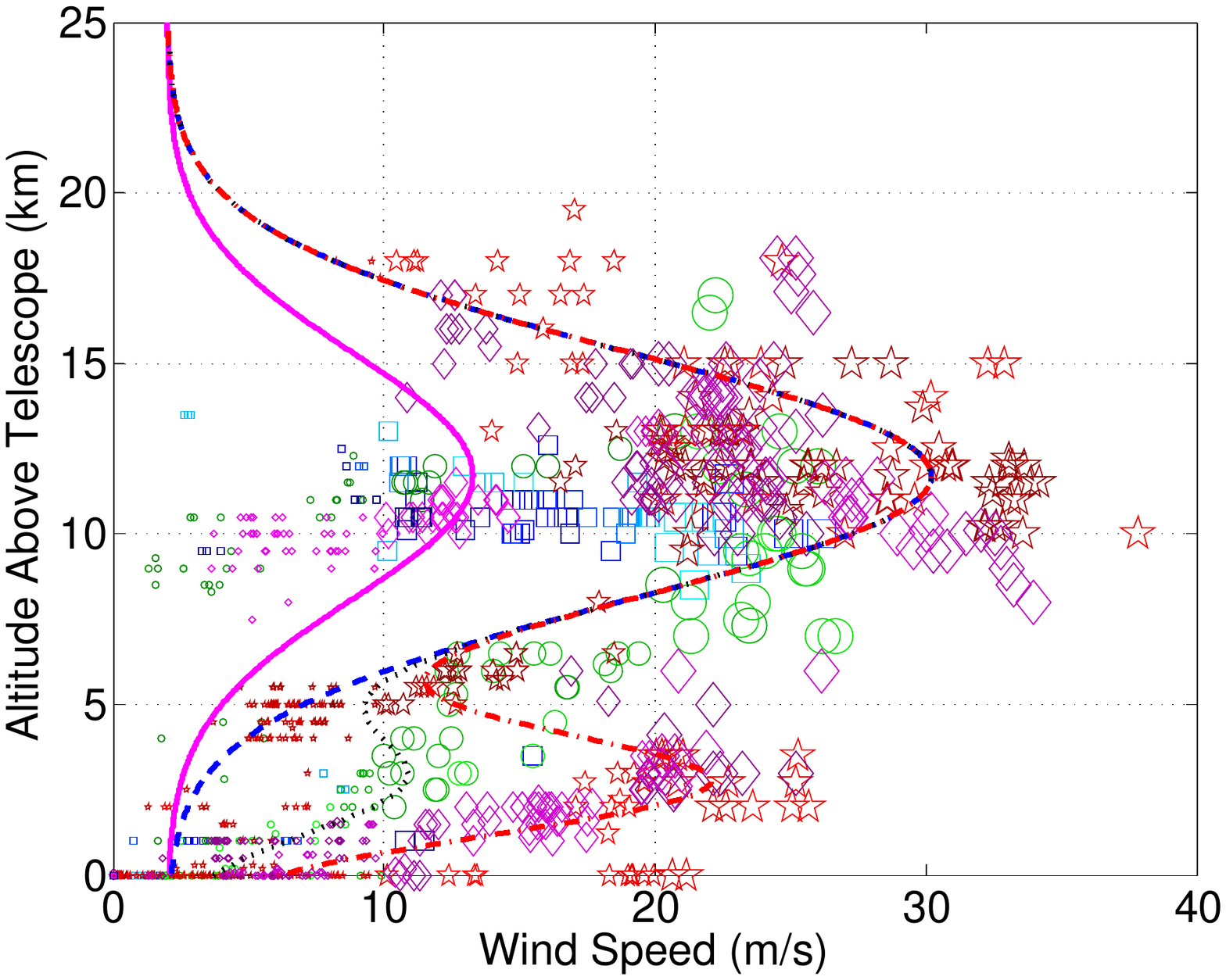}\label{fig:trending:all_vel_Model:fit}}
  \subfigure[MJUO \vw\s models]{\includegraphics[width=0.8\linewidth, trim=0mm 70mm 0mm 90mm]{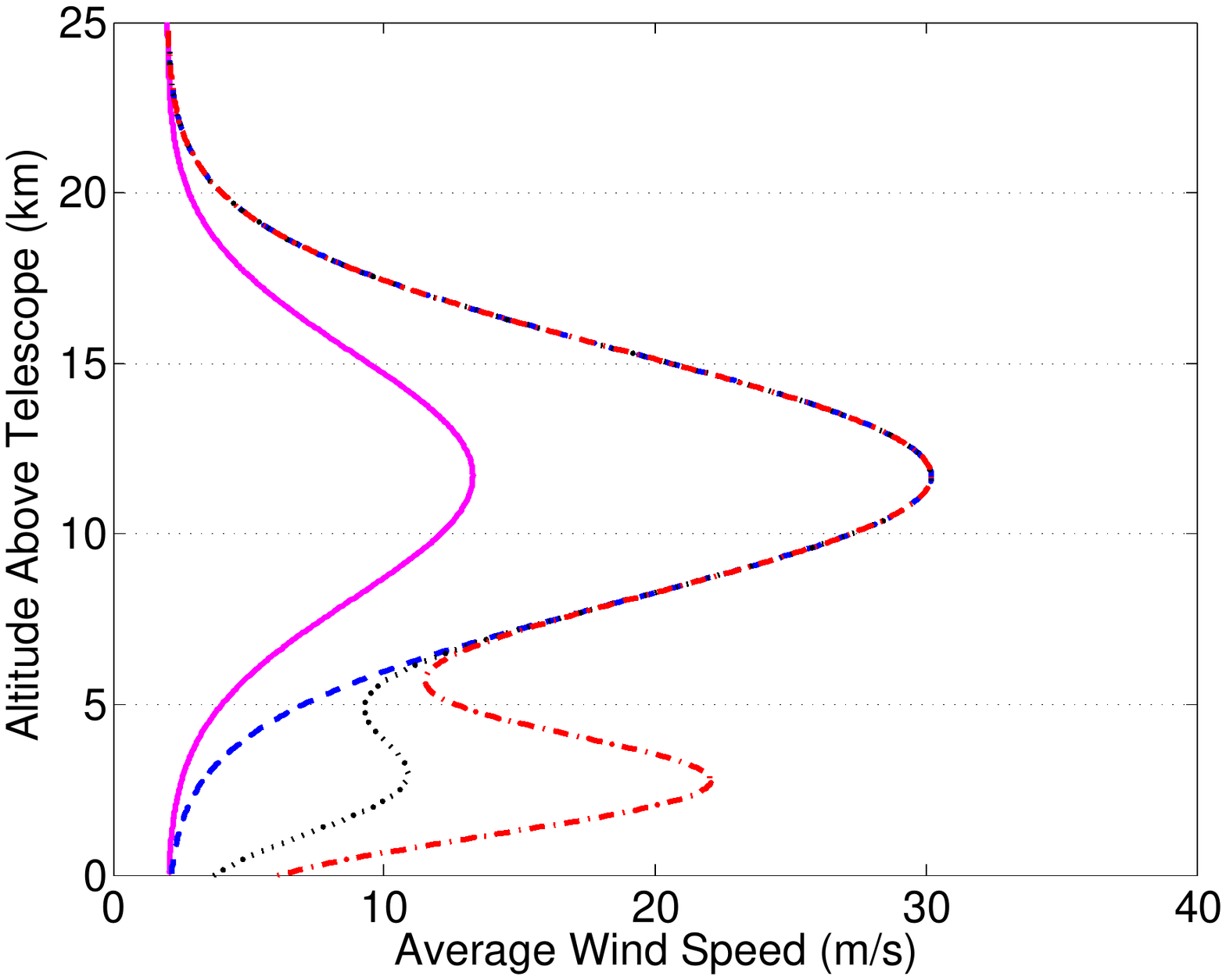}\label{fig:trending:all_vel_Model:models}}\\
  \caption[Fit of \vw\s models to measured profiles.]{Fit of \vw\s models to measured profiles. Models shown are (\textcolor{magenta}{---})
  MJUO1V, (\textcolor{blue}{-- --}) MJUO2V, ($\cdots$) MJUO3V and (\textcolor{red}{-- $\cdot$ --}) MJUO4V.}\label{fig:trending:all_vel_Model}
\end{figure}

Unlike the modelling of \cn, it is difficult to employ a generic model that would encompass the majority of conditions, as the velocity seen in the upper layers is dependent on the velocity detected near the ground.  Figure \ref{fig:trending:all_vel_Model:fit} shows  the instantaneous wind speeds obtained during June and July of 2005 and May and June of 2007.  Overlaid are four different \vw\s models developed to encompass the range of velocity characteristics detected.  The parameters for the four models for MJUO are listed in Table \ref{tab:trending:VelModelParams}.

MJUO1V (indicated by the solid purple line in Figures \ref{fig:trending:all_vel_Model:fit} and \ref{fig:trending:all_vel_Model:models}) is designed for very calm and clear nights where little NGT is present and the high altitude turbulence has low wind speeds.  MJUO2V (indicated by dashed black line) was developed for nights such as 2007 May 3,  where high altitude turbulence is strong, but low-altitude layer wind speeds are very low.  Both MJUO1V and MJUO2V are based on the traditional Greenwood wind model.  MJUO3V (indicated by the solid red line) and MJUO4V (indicated by the dashed blue line) employ a second Gaussian peak described in equation (\ref{eqn:trending:ModifiedGreenwoodVelocity}).  MJUO3V is intended for moderate ground wind speeds measuring 2.8 -- 5.6 \vhunits.  Based on UC-SCIDAR measurements obtained this is the most likely situation to be encountered and hence should be the preferred model for AO design.  MJUO4V was developed for situations where high ground wind speeds are present.



\begin{table*}
\begin{minipage}[c]{\linewidth}
\centering
\caption{Parameters for wind velocity \vw\s models.}\label{tab:trending:VelModelParams}
\small{\begin{center}\begin{tabular}{ccccccccccc}
\hline
 Model &  Usage\footnote{Usage is indicative of the ground wind velocity conditions: \newline calm: 0 \vhunits; low: $< 2.8$ \vhunits; mod: 2.8 -- 5.6 \vhunits; high: $> 5.6$ \vhunits} &  $V(0)$ &  $V(H_T)$ &  $H_T$ &  $L_T$ &  $V(H_1)$ &  $H_1$ &  $L_1$ &  $\fg($HV 5-7$)$\footnote{\fg\s values are specified for $\lambda = 589$ nm.} & $\fg($MJUO3$)$ \\
  &   &  (\vhunits) &  (\vhunits) &  (km) &  (km) &  (\vhunits) &  (km) &  (km) &  (Hz) & (Hz) \\
\hline
Bufton &  --- &  5 &  30 &  9.8 &  4.8 &  0 &   &   &  29.5 & 40.6 \\
 MJUO1V &  calm &  2 &  12 &  11 &  4.8 &  0 &   &   &  19.9 & 32.0 \\
 MJUO2V &  low  &  2 &  30 &  11 &  4.8 &  0 &   &   &  36.5 & 63.6 \\
 MJUO3V &  mod &  2 &  30 &  11 &  4.8 &  8 &  2.5 &  2 &  46.4 & 76.9 \\
 MJUO4V &  high &  2 &  30 &  11 &  4.8 &  20 &  2.5 &  2 &  65.9 & 105.0 \\
\hline
\end{tabular}\end{center}}
\end{minipage}
\end{table*}

Using the MJUO3 \cn\s model and the MJUO3V \vw\s model for moderate ground wind speeds, \fg\s is estimated to be 79 Hz for a wavelength of 589 nm, which is more in line with the 30 -- 90 Hz range suggested by \citet{TysonB48}.

  Although adequate models for \vh\s profiles were determined, it is recommended that more temporal data be collected to refine the models.  An investigation into the correlation between low and high altitude velocities could also be conducted.

\section{Conclusions}
\label{sec:conclusions}

UC-SCIDAR measurements taken between 2005 -- 2007 detected strong NGT with a weaker layer located at 12 -- 14 km above sea level.  On calm nights a third mid-altitude layer was detected at $\sim6$ km above sea level.  In a significant amount of data strong low altitude turbulence extended up to 5 km above sea level.  The measurements suggest an average $r_0$ of $12\pm5$ cm and $7\pm1$ cm for \pup\s and \gen\s profiles respectively.  This corresponds to an angular resolution, $\theta_{\mathrm{res}}$, of 2.1 arcsec for the full profile.  The average $\theta_0$ values were $1.5\pm0.5$ arcsec and $1.0\pm0.1$ arcsec for \pup\s and \gen\s profiles respectively.  Average $\overline{h_0}$ values were $6\pm1$ km and $2.0\pm0.7$ km for \pup\s and \gen\s measurements respectively.

During spatial analysis, correction factors were applied to generalised profiles to the layers detected in the free atmosphere in the post-processing phase.  It would be preferable to have these corrections incorporated into the inversion algorithm to reduce the amount of double handling of data.  To create an automatic detection system for this, a thorough investigation as to the effects of NGT strength on the detected strength from higher layers should be conducted, including various altitudes and high altitude turbulence strength.

Temporal analysis detected layers located at similar altitudes, with tropopause layer velocities of 12 -- 30 \vhunits, dependent on weather conditions.  Low altitude turbulence layers had velocities ranging from 2 \vhunits\s to well over 24 \vhunits.  No trends could be established for the values of \fg, due to the gaps in the \vh\s profiles.  Little seasonal variation was detected in the \cn\s profiles.  Both \cn\s and \vw\s profiles were highly dependent on the weather conditions.

A modified Hufnagel-Valley (HV) model was developed to describe the \cn\s profiles, incorporating a strong NGT layer, a layer at 11 km above the telescope (i.e. 12 km above sea level) and two additional layers: one at 5.5 km above the telescope and the other at 1.5 km above the telescope extending up to 4 km.  The resulting model estimates an $r_0$ of 6 cm for a wavelength of 589 nm, which corresponds to a $\theta_{\mathrm{res}}$ of 2.5 arcsec.  $\theta_0$ is estimated at 0.9 arcsec.

A series of \vw\s models were developed, based on the Greenwood wind model with an additional Gaussian peak located at low altitudes to model the \vh\s profiles seen at MJUO.  The models correspond to calm, light, moderate and strong ground wind speed conditions seen at the site.  Using the modified HV model for \cn\s profiles and the suggested model for \vh\s profiles in the presence of moderate ground wind speeds, \fg\s was estimated at 79 Hz for a wavelength of 589 nm.


\section*{Acknowledgments} 
The authors would like to thank C. Clare Worley (former student at University of Canterbury, now at Observatoire de la Cote d'Azur) for collecting the data during 2005.  A significant portion of this research used this data.  Thank you to Steve Weddell (University of Canterbury) for the loan of the cameras used in the current UC-SCIDAR system.  J.L. Mohr also acknowledges Dr Charles Jenkins and Dr Andrew Lambert for comments on her PhD thesis in which this data was presented.



\begin{thebibliography}{}

\bibitem[{{Andrews}}(2004)]{AndrewsB47}{Andrews}, L.~C. 2004, {{Field Guide to Atmospheric Optics}}, vol. FG03 of {SPIE
Field Guides}. SPIE Press, Bellingham, WA

\bibitem[{{Avila} et~al.}(1997)]{AvilaJ138}{Avila}, R., {Vernin}, J. \& {S{\' a}nchez}, L.~J. 1997, {\ao}, {36},
7898

\bibitem[{{Avila} et~al.}(2001)]{AvilaJ149} Avila, R., Vernin, J. \& Cuevas, S. 1998, {\pasp}, {110}, 1106

\bibitem[{{Avila} et~al.}(2001)]{AvilaJ65}{Avila}, R., {Vernin}, J. \&  {S{\' a}nchez}, L.~J. 2001, {\aap}, {369}, 364

\bibitem[{{Avila} et~al.}(2004)]{AvilaJ267} {Avila}, R., {Masciadri}, E., {Vernin}, J. \& S{\'{a}}nchez, L.~J. 2004, {\pasp}, {116}, 682

\bibitem[{{Avila} et~al.}(2007)]{AvilaJ326} Avila, R., Sanchez, L. J., Iba\~{n}ez, F., Masciadri, E., Azouit, M.; Agabi, A. \& Cuevas, S. 2007, RevMexAA (Serie de Conferencias), 31, 71

\bibitem[{{Avila} et~al.}(2008)]{AvilaMNRAS2008} {Avila}, R., {Avil{\'e}s}, J.~L., {Wilson}, R.~W., {Chun},
	M., {Butterley}, T. \& {Carrasco}, E. 2008, {\mnras}, {387}, 1511

\bibitem[{{Fried}}(1966)]{FriedJ132}{Fried}, D.~L. 1966, {\josa}, { 56}, 1372

\bibitem[{{Fuensalida} et~al.}(2004)]{FuensalidaC89}{Fuensalida}, J.~J., et~al. 2004, in {{Advancements in Adaptive Optics}}, ed. D.~B. {Calia}, B.~L. {Ellerbroek}, and R.~{Ragazzoni}, { \spie}, 5490, 749


\bibitem[{{Garc{\'{i}}a-Lorenzo} et~al.}(2009)]{GarciaLorenzoMNRAS2009} {Garc{\'{i}}a-Lorenzo}, B., Eff-Darwich, A., {Fuensalida}, J.~J., \& Castro-Almazan, J. 2009, \mnras, 397, 1633

\bibitem[{Hardy}(1998)]{HardyBook1998}Hardy, J.~W. 1998, {Adaptive Optics for Astronomical Telescopes}, no.~16 in Oxford Series in Optical and Imaging Sciences. Oxford University Press, New York

\bibitem[Hearnshaw et~al. (2002)]{HearnshawExA2002}{Hearnshaw}, J.~B., {Barnes}, S.~I., {Kershaw}, G.~M., {Frost},
	N., {Graham}, G., {Ritchie}, R. \& {Nankivell}, G.~R. 2002, ExA, {13}, 59

\bibitem[Johnston (2000)]{JohnstonP2} Johnston, R. A. 2000, Inverse Problems in Astronomical Imaging, PhD Thesis, University of Canterbury

\bibitem[{{Johnston} et~al.}(2000)]{JohnstonJ200}{Johnston}, R.~A., {Connolly}, T.~J. \& {Lane}, R.~G. 2000, {\optcomm}, {181}, 267

\bibitem[{Johnston \& Lane}(2000)]{JohnstonJ236}{Johnston}, R.~A. \& {Lane}, R.~G. 2000, {\ao}, {39}, 4761

\bibitem[{{Johnston} et~al.}(2002)]{JohnstonJ44}{Johnston}, R.~A., {Dainty}, C., {Wooder}, N.~J. \&  {Lane}, R.~G. 2002, {\ao}, {41}, 6768

\bibitem[{{Johnston} et~al.}(2004)]{JohnstonC67}{Johnston}, R.~A., {Mohr}, J.~L., {Cottrell}, P.~L. \& {Lane}, R.~G.  2004, in {{Proceedings of Image and Vision Computing New Zealand 2004}}, ed. D.~{Pairman}, H.~{North}, and S.~{McNeill}, 209

\bibitem[{{Johnston} et~al.}(2005)]{JohnstonC71}{Johnston}, R.~A., {Worley}, C.~C.,  {Mohr}, J.~L.,  {Lane}, R.~G. \&  {Cottrell}, P.~L. 2005, in {{Proceedings of Image and Vision Computing New Zealand 2005}}, ed. B.~{McCane}, 487

\bibitem[{{Kl{\"{u}}ckers} et~al.}(1998)]{KluckersJ67}{Kl{\"{u}}ckers}, V.~A., {Wooder}, N.~J., {Nicholls}, T.~W., {Adcock}, M.~J., {Munro}, I.~H. \&  {Dainty}, J.~C. 1998, { \aaps}, { 130}, 141

\bibitem[{Masciadri et~al.}(2010)]{MasciadriMNRAS2010} Masciadri, E., Stoesz, J., Hagelin, S. \& Lascaux, F. 2010, \mnras, in press

\bibitem[{{Mohr} et~al.}(2006)]{MohrC84}{Mohr}, J.~L.,  {Johnston}, R.~A., {Worley}, C.~C.  \&  {Cottrell}, P.~L. 2006, in {{Proceedings of Image and Vision Computing New Zealand 2006}}, ed. P.~{Delmas}, J.~{James}, and J.~{Morris}, 523

\bibitem[{Mohr et~al.}(2008a)]{MohrIVCNZ08}Mohr, J.~L., Johnston, R.~A. \&  Cottrell, P.~L. 2008a, in {23rd International Conference Image and Vision Computing New Zealand 2008}, ed. K.~Irie, and D.~Pairman, DOI:10.1109/IVCNZ.2008.4762089

\bibitem[{{Mohr} et~al.}(2008b)]{MohrSPIE2008}{Mohr}, J.~L., {Johnston},R.~A., {Worley}, C.~C. \& {Cottrell}, P.~L. 2008b, in {Optics in Atmospheric Propagation and Adaptive Systems XI}, ed. A.~Kohnle, K.~Stein, and J.~D. Gonglewski, { \spie}, 7108, 710809

\bibitem[Mohr (2009)]{MohrPhD2009} Mohr, J. L. 2009, Atmospheric Turbulence Characterisation Using Scintillation Detection and Ranging, PhD Thesis, University of Canterbury

\bibitem[{{Northcott}}(1999)]{NorthcottAB5}{Northcott}, M.~J. 1999, in {{Adaptive Optics in Astronomy}}, ed. F.~{Roddier},
Cambridge University Press, Cambridge, 155

\bibitem[{{Parenti} and {Sasiela}}(1994)]{ParentiJ100}{Parenti}, R.~R. \&  {Sasiela}, R.~J. 1994, { \josaa}, { 11}, 288

\bibitem[{{Prieur} et~al.}(2001)]{PrieurJ58}{Prieur}, J.-L., {Daigne}, G. \& {Avila}, R. 2001, { \aap}, { 371}, 366

\bibitem[{{Roddier}}(1981)]{RoddierAB12}{Roddier}, F. 1981, in {{Progress in Optics}}, ed. E.~{Wolf}, North-Holland
Publishing Company, Amsterdam, Netherlands, 19, 281

\bibitem[{{Sturman} and {Tapper}}(1996)]{SturmanB49}{Sturman}, A. \& {Tapper}, N. 1996, {{The Weather and Climate of Australia and New Zealand}}. Oxford University Press, Melbourne, Australia

\bibitem[{{Tokovinin} et~al.}(2005)]{TokovininJ333}{Tokovinin}, A., {Vernin}, J., {Ziad}, A. \& {Chun}, M.~R. 2005, {\pasp}, { 117}, 395

\bibitem[{{Tyson}}(1991)]{TysonB42}{Tyson}, R.~K. 1991, {{Principles of Adaptive Optics}}. Academic Press, Inc., Boston

\bibitem[{{Tyson} \& {Frazier}}(2004)]{TysonB48}{Tyson}, R.~K. \& {Frazier}, B.~W. 2004, { {Field Guide to Adaptive Optics}}, vol. FG03 of { SPIE Field Guides}. SPIE Press, Bellingham, WA

\bibitem[{{Wang} et~al.}(2008)]{WangJ330}{Wang}, L., {Sch{\" o}ck}, M. \& {Chanan}, G. 2008, { \ao}, { 47}, 1880

\end{thebibliography}
\end{document}